\documentclass[10pt,a4paper]{article}
\usepackage{amsmath}
\usepackage{amsfonts}
\usepackage{amssymb}
\usepackage[dvips]{graphicx}

\begin{document}

\title{\textbf{Nucleation of vacuum bubbles in\\ Brans-Dicke type theory}\footnote{CQUeST-2011-0404}}
\author{\textsc{Hongsu Kim}$^{a}$\footnote{hongsu@kasi.re.kr}\;, \textsc{Bum-Hoon Lee}$^{b,c}$\footnote{bhl@sogang.ac.kr}\;, \textsc{Wonwoo Lee}$^{c}$\footnote{warrior@sogang.ac.kr}\;,\\
\textsc{Young Jae Lee}$^{d}$\footnote{noasac@hotmail.com}\; \textrm{and} \textsc{Dong-han Yeom}$^{c,d,e}$\footnote{innocent.yeom@gmail.com}\\
$^{a}$\small{\textit{Korea Astronomy and Space Science Institute, Daejeon 305-348, Republic of Korea}}\\
$^{b}$\small{\textit{Department of Physics, BK21 Division, Sogang University, Seoul 121-742, Republic of Korea}}\\
$^{c}$\small{\textit{Center for Quantum Spacetime, Sogang University, Seoul 121-742, Republic of Korea}}\\
$^{d}$\small{\textit{Department of Physics, KAIST, Daejeon 305-701, Republic of Korea}}\\
$^{e}$\small{\textit{Research Institute for Basic Science, Sogang University, Seoul 121-742, Republic of Korea}}
}

\maketitle
\begin{abstract}
In this paper, we explore the nucleation of vacuum bubbles in the Brans-Dicke type theory of gravity.
In the Euclidean signature, we evaluate the fields at the vacuum bubbles as solutions of the Euler-Lagrange equations of motion as well as the bubble nucleation probabilities by integrating the Euclidean action.
We illustrate three possible ways to obtain vacuum bubbles: true vacuum bubbles for $\omega > -3/2$, false vacuum bubbles for $\omega < -3/2$,
and false vacuum bubbles for $\omega > -3/2$ when the vacuum energy of the false vacuum in the potential of the Einstein frame is \textit{less} than that of the true vacuum.
After the bubble is nucleated at the $t=0$ surface, we can smoothly interpolate the field combinations to some solutions in the Lorentzian signature and consistently continue their subsequent evolutions.
Therefore, we conclude that, in general scalar-tensor theories like this Brans-Dicke type theories, which may include and represent certain features of string theory, vacuum bubbles come in false vacuum bubbles as well as in true vacuum bubbles, as long as a special condition is assumed on the potential.
\end{abstract}

\newpage

\tableofcontents

\newpage

\newpage

\section{Introduction}

In this paper, we study (Euclidean) vacuum bubble nucleations and their subsequent evolution in the context of the Brans-Dicke type theory of gravity \cite{brdi01}.

The Brans-Dicke theory \cite{brdi01} is the most studied and hence the best known among all the alternative theories of classical gravity to Einstein's general relativity \cite{will}.
Historically, this theory has been thought of as a minimal extension of general relativity that properly accommodates both Mach's principle and Dirac's large number hypothesis \cite{will}\cite{weinberg}.
The action of the Brans-Dicke theory takes the form
\begin{equation}
S_{\mathrm{E}}[g, \Phi] = \frac{1}{16 \pi} \int d^{4}x \sqrt{-g} \left( \Phi R - \frac{\omega}{\Phi} \Phi_{;\mu}\Phi_{;\nu}g^{\mu\nu} \right),
\end{equation}
where $R$ is the Ricci scalar, $\omega$ is a dimensionless coupling parameter, and $\Phi$ is the Brans-Dicke scalar field.
The theory employs the viewpoint that Newton's constant $G$ is allowed to vary with space and time and can be written in terms of a scalar field $\Phi$ as $G = 1/\Phi$.
As a scalar-tensor theory of gravity, the Brans-Dicke theory involves an adjustable but undetermined Brans-Dicke parameter $\omega$. As is well known, the larger the value of $\omega$, the more dominant the tensor (curvature) degree, and the smaller the value of $\omega$, the larger the effect of the Brans-Dicke scalar.
As long as we select a sufficiently large value of $\omega$, the prediction of the theory will agree with all observations/experiments \cite{will}.
For this reason, the Brans-Dicke theory has remained a viable theory of gravity.
Moreover, interesting models \cite{hongsu} that explain dark matter and dark energy have been developed, possibly implying that the Brans-Dicke theory may be a more relevant theory of classical gravity that is consistent with observations.

In this paper, we focus on nucleations of vacuum bubbles. It is thus convenient to use the Euclidean signature. Generally speaking, in non-linear field theories, there are non-topological soliton configurations. These are solutions of classical field equations in pure scalar field theories with non-linear potential terms. An interesting and significant example for such a non-topological soliton configuration is the true vacuum bubble. That is, the bubble arises via quantum tunneling (i.e., the super cooled first-order cosmological phase transition) from the high temperature symmetric false vacuum state to the low temperature symmetry-breaking true vacuum state. Along this line, the dynamics of quantum tunneling was first developed by \cite{Coleman:1980aw} in the flat space-time background and by \cite{CDL}\cite{parke} in the curved space-time background. The formulation that we shall employ in the present work can indeed be regarded as an extension or generalization of this last reference.

In order to study the nucleation and evolution of vacuum bubbles, we need a non-linear potential that can give metastable local false vacua and a global true vacuum. One possible way is to employ the Brans-Dicke gravity which involves a scalar field with such a potential. If this is possible, it will be an interesting model since it preserves the weak equivalence principle, which was the initial motivation of the Brans-Dicke theory. In this paper, however, as a toy model, we shall introduce a potential in the Brans-Dicke field sector. Indeed, if we relax the original constraints of Brans and Dicke that protect the weak equivalence principle, the non-linear potential for the Brans-Dicke scalar field can be used and it would allow vacuum bubble solutions. And this is why we call this kind of theory a Brans-Dicke \textit{type} theory.

To summarize our results in advance, we illustrate three possible ways to obtain vacuum bubbles: true vacuum bubbles for $\omega > -3/2$, false vacuum bubbles for $\omega < -3/2$, and false vacuum bubbles for $\omega > -3/2$ when the vacuum energy of the false vacuum in the potential of the Einstein frame is \textit{less} than that of the true vacuum. (The third solution of a false vacuum bubble is related to the authors' previous papers due to a non-minimally coupled field \cite{lll2006}.) After the bubble is nucleated at the $t=0$ surface, we can smoothly interpolate the field combinations to the solutions in the Lorentzian signature and consistently continue their subsequent evolutions. Note that, in this paper, we evaluate field configurations and probability amplitudes in the Jordan frame in a consistent manner.

This paper is organized as follows. In Section~\ref{sec:Euc}, we discuss the Euclidean action of the Brans-Dicke type theory. In Section~\ref{sec:Nuc}, we classify and confirm possible nucleation processes of vacuum bubbles in the Brans-Dicke type theory. In Section~\ref{sec:Nucandevo}, we discuss nucleation and evolution of false vacuum bubbles in the thin wall approximation. Finally, in Section~\ref{sec:Dis}, we summarize the present study and discuss related problems.

\section{\label{sec:Euc}Euclidean action in Brans-Dicke type theory}

In this section, we describe the Brans-Dicke type theory in the Euclidean signature.

\subsection{Brans-Dicke type theory in the Euclidean signature}

The action of the Brans-Dicke type theory \cite{brdi01} with a potential takes the following form:
\begin{eqnarray}
S_{\,\mathrm{E}} = \int\sqrt{g}d^{4}x \; \mathcal{L}_{\mathrm{BD}},
\end{eqnarray}
where the Lagrangian density is
\begin{eqnarray}
\mathcal{L}_{\mathrm{BD}} = \frac{1}{16\pi} \left( - \Phi R + \frac{\omega}{\Phi}\Phi_{;\mu}\Phi_{;\nu}g^{\mu\nu} + V(\Phi) \right).
\end{eqnarray}
Here, $\sqrt{g}=\sqrt{+\det g}$, $\Phi$ is the Brans-Dicke scalar field, $R$ is the Ricci scalar, $\omega$ is the dimensionless coupling parameter of the Brans-Dicke type theory, and $V(\Phi)$ is the potential of the Brans-Dicke scalar field.

By extremizing this action with respect to the metric $g_{\mu\nu}$ and the Brans-Dicke field $\Phi$, one gets the classical Euler-Lagrange equations of motion given, respectively, by \cite{Faraoni:2004pi}
\begin{eqnarray}
G_{\mu\nu} = \frac{1}{\Phi} \left(-g_{\mu\nu}\Phi_{;\rho \sigma}g^{\rho\sigma}+\Phi_{;\mu\nu}\right) + \frac{\omega}{\Phi^{2}} \left(\Phi_{;\mu}\Phi_{;\nu}-\frac{1}{2}g_{\mu\nu}\Phi_{;\rho}\Phi_{;\sigma}g^{\rho\sigma}\right) -g_{\mu\nu}\frac{V(\Phi)}{2\Phi},
\end{eqnarray}
\begin{eqnarray}
\Phi_{;\mu\nu}g^{\mu\nu}= \frac{1}{3+2\omega}\left(\Phi V'(\Phi) - 2V(\Phi)\right).
\end{eqnarray}

In the following subsections, we discuss the possible origin of our choices of the coupling $\omega$ and the potential $V(\Phi)$. In this paper, we work in the geometrical unit $c=G=1$.

\subsubsection{Where does $\omega$ come from?}

From observational tests, it is known that the value of $\omega$ should be greater than $4 \times 10^{4}$ \cite{Ber}. However, in various physical models, small $\omega$ parameters can be allowed. Even though a small $\omega$ is not for our Universe, if a small $\omega$ is allowed in the fundamental theory and if such a small value of $\omega$ has implications, the study of various $\omega$ will have theoretical importance.

We now start with the example of dilaton gravity, which has the effective action in the following form \cite{Gasperini:2007zz}:
\begin{eqnarray}
\label{eq:dilaton} S = \frac{1}{2 \lambda_{s}^{d-1}}\int d^{d+1}x \sqrt{-g} e^{-\phi} \left( R + (\nabla \phi)^{2} \right),
\end{eqnarray}
where $d$ is the space dimensions, $\lambda_{s}$ is the length scale of string units, $R$ is the Ricci scalar, and $\phi$ is the dilaton field. It is interesting to note that a simple field redefinition brings this dilaton gravity into a Brans-Dicke type theory. That is, if we define $\Phi$ as
\begin{eqnarray}
\label{eq:def} \frac{e^{-\phi}}{\lambda_{s}^{d-1}} = \frac{\Phi}{8 \pi G_{d+1}},
\end{eqnarray}
where $G_{d+1}$ is the $d+1$-dimensional gravitation constant, then we end up with a Brans-Dicke type theory with $\omega = -1$.

If there are higher loop corrections coming from string theory, there will be other coupling terms in $\phi$. For example, the effective action of heterotic string theory compactified on a $Z_{N}$ orbifold takes the following form \cite{Foffa:1999dv}:
\begin{eqnarray}
S = \frac{1}{2 \lambda_{s}^{2}}\int d^{4}x \sqrt{-g} e^{-\phi} \left( R + (1+e^{\phi}G(\phi))(\nabla \phi)^{2} \right),
\end{eqnarray}
where
\begin{eqnarray}
G(\phi)= \left( \frac{3\kappa}{2} \right) \frac{6+\kappa e^{\phi}}{(3+\kappa e^{\phi})^{2}}
\end{eqnarray}
and $\kappa$ is a positive constant of order one which is determined by the coefficients of the anomaly. The coupling parameter should then be field dependent: $\omega(\Phi) = -1-e^{\phi}G(\phi)$. In this specific model, $\omega$ depends on $\lambda_{s}$ and $\kappa$, and it is possible to find $\omega < -3/2$.
For the case at hand, obviously, the Brans-Dicke scalar field equation is subject to change and it turns out to be \cite{Faraoni:2004pi}
\begin{eqnarray}
\Phi_{;\mu\nu}g^{\mu\nu}= \frac{1}{3+2\omega}\left(\Phi V'(\Phi) - 2V(\Phi) -\frac{d\omega}{d\Phi} \Phi_{;\mu}\Phi_{;\nu}g^{\mu\nu} \right).
\end{eqnarray}
Therefore, if the variation in $\Phi$ is sufficiently small, and hence the variation of $\omega(\Phi)$ is sufficiently small, an $\omega$ of less than $-3/2$ could be naturally justified.

In the first model of Randall and Sundrum \cite{Randall:1999ee}, two branes have been employed to account for the hierarchy problem. Because of the warp factor between the two branes, we obtain a positive tension brane and a negative tension brane in the anti-de Sitter (AdS) space background. According to Garriga and Tanaka \cite{Garriga:1999yh}, it is interesting to note that, each brane can be described by the Brans-Dicke type theory in the weak field limit with the $\omega$ parameter
\begin{eqnarray}
\label{eq:gt} \omega = \frac{3}{2} \left( e^{\pm s/l} - 1 \right),
\end{eqnarray}
where $s$ is the location of the negative tension brane along the fifth dimension, $l=\sqrt{-6/\Lambda}$ is the length scale of the anti-de Sitter space, and the sign $\pm$ denotes the sign of the tension. To explain the hierarchy problem, we require $s/l \sim 35$. We then obtain a sufficiently large value of $\omega$ on the positive tension brane while $\omega \gtrsim -3/2$ on the negative tension brane \cite{Garriga:1999yh}\cite{Fujii:2003pa}. In principle, however, $s/l$ can be chosen arbitrarily, and hence one may infer that various $\omega$ near $-3/2$ may be allowed by models of the brane world scenarios.

\subsubsection{Comments on the choice of the Jordan frame}

In the present work, we consistently calculate all quantities in the Jordan frame. In this subsection, we briefly comment on the choice of the Jordan frame and its merits.

First of all, the conformal transformation from the Jordan frame to the Einstein frame is consistent only if $\omega > -3/2$. Therefore, in general Brans-Dicke type models, the Jordan frame is more general than the Einstein frame, in some sense.

If $\omega > -3/2$, then it is possible to transform a solution of a vacuum bubble from the Jordan frame to the Einstein frame. However, it is not trivial whether the bubble in the Einstein frame is Coleman-De Luccia type \cite{CDL}\cite{parke} or some other type \cite{LL}\cite{LW}; also it is not trivial whether it is a true vacuum bubble or a false vacuum bubble.

In previous work \cite{Lee:2010yd}, the authors studied the dynamics of thin wall bubbles for Brans-Dicke type theories in Lorentzian signatures. One of the interesting results is that a thin wall of a false vacuum bubble may violate the null energy condition in the Jordan frame. However, since we assumed the thin wall approximation, such a property seemed not to depend on the choice of $\omega$. If a bubble expands along a causal patch in the Jordan frame, it should be the same in the Einstein frame. However, we already know that if $\omega > -3/2$, the Einstein frame does not violate the null energy condition, and it may imply that such dynamics is not consistent in the Einstein frame. Then, the natural open question is, what is the proper interpretation of such bubbles in the Einstein frame for a given $\omega$? Is it consistent even if we do not assume the thin wall approximation?

To answer this question, the natural direction of study is, first, to classify possible small false vacuum bubbles in the Jordan frame, and second, to interpret the meanings of such bubbles in the Einstein frame. Accordingly, in this paper, we choose the Jordan frame to study a nucleation of vacuum bubbles.

\subsubsection{Effective potential: a toy model}

Generally, string theory predicts a non-minimal and non-universal coupling of various fields to the dilaton \cite{Gasperini:2007zz}. Hence, we may well include the potential term for the dilaton field in the original action, if we pay the price of violating the weak equivalence principle.

In practice, obviously, we need to choose a specific potential. In this paper, the origin of such a potential is not our concern, and thus we will not address this issue. Rather, we will take a simple potential and explore its consequences.

We now start with the effective force function $F(\Phi)$ given by
\begin{eqnarray}
F(\Phi) &\equiv& \Phi V'(\Phi) - 2 V(\Phi) \\
&=& A \left(\Phi-\Phi_{\mathrm{t}} \right) \left(\Phi-\Phi_{\mathrm{f}} \right) \left(\Phi- \left( \frac{\Phi_{\mathrm{t}}+\Phi_{\mathrm{f}}}{2} + \delta \right) \right),
\end{eqnarray}
where $A$ is a positive constant, $\Phi_{\mathrm{t}}$ and $\Phi_{\mathrm{f}}$ denote the field value inside or outside the vacuum bubble (the subscript $\mathrm{t}$ denotes a true vacuum and the subscript $\mathrm{f}$ denotes a false vacuum), and $\delta$ is a free parameter that determines the location of the bump of the potential. We can then choose that the inside and outside regions to be in stable equilibrium.

In the present work, for convenience, we choose $\Phi_{\mathrm{t}}=1$ and $V(\Phi_{\mathrm{t}})=V_{0}$ in the true vacuum region. The potential $V(\Phi)$ and the effective potential $U(\Phi)$ then take the following form:
\begin{eqnarray}
V(\Phi) = \Phi^{2} \left( \int_{1}^{\Phi} \frac{F(\bar{\Phi})}{\bar{\Phi}^{3}} d \bar{\Phi} + V_{0} \right)
\end{eqnarray}
and
\begin{eqnarray}
U(\Phi) = \int_{1}^{\Phi} F(\bar{\Phi}) d \bar{\Phi} = \int_{1}^{\Phi}\left( \bar{\Phi} V'(\bar{\Phi}) - 2V(\bar{\Phi}) \right) d \bar{\Phi},
\end{eqnarray}
with the field equation being given by $\nabla^{2} \Phi = U'/(3+2\omega)$.

Note that the potential in the Einstein frame $U_{E}$ is \cite{Faraoni:2004pi}
\begin{eqnarray}
U_{E}(\Phi) = \int_{1}^{\Phi} \frac{F(\bar{\Phi})}{\bar{\Phi}^{3}} d \bar{\Phi} + V_{0}.
\end{eqnarray}
Of course, we have to represent $U_{E}$ by a new field $\Phi_{E}$, where
\begin{eqnarray}
\Phi = \exp \Phi_{E}\sqrt{\frac{16 \pi}{2\omega+3}},
\end{eqnarray}
in order to make the canonical action in the Einstein frame. However, the relation between $\Phi$ and $\Phi_{E}$ is one-to-one and onto. Therefore, the only effect is to stretch the potential along the field direction, and this does not affect the vacuum energy at each field value.

\subsection{Euclidean action in Brans-Dicke type theory}

Now we evaluate the Euclidean action of the Brans-Dicke type theory to calculate the probability amplitude of bounces. First, as usual, we assume the $O(4)$ symmetric metric \cite{Coleman:1980aw}\cite{CDL}\cite{parke}\cite{LW}\cite{LL}:
\begin{eqnarray}
ds^{2}_{\mathrm{E}} = d\eta^{2} + \rho^{2}(\eta) (d\chi^{2} + \sin^{2} \chi (d\theta^{2} + \sin^{2}\theta d\varphi^{2})),
\end{eqnarray}
where $\rho$ is a function that corresponds to the scale factor in the Lorentzian signature, $\eta$ is the Euclidean time parameter, and $\chi$, $\theta$, and $\varphi$ are angle coordinates on the three-dimensional sphere.

In terms of this $O(4)$ symmetric metric, then, the classical Euler-Lagrange equations of motion are given by
\begin{eqnarray}
G_{\eta\eta} &=& 3 \frac{\dot{\rho}^{2}-1}{\rho^{2}} = -3 \frac{\dot{\rho}}{\rho} \frac{\dot{\Phi}}{\Phi} + \frac{\omega}{2} \left(\frac{\dot{\Phi}}{\Phi}\right)^{2} - \frac{V}{2\Phi}
\end{eqnarray}
and
\begin{eqnarray}
\nabla^{2} \Phi &=& \ddot{\Phi} + 3 \frac{\dot{\rho}}{\rho} \dot{\Phi} \\
&=& \frac{1}{2 \omega + 3} \left(\Phi V'(\Phi) - 2V(\Phi)\right),
\end{eqnarray}
where the over-dot denotes a derivative with respect to $\eta$.

Note that the two key equations to evaluate the Euclidean action of the Brans-Dicke type theory are
\begin{eqnarray}
(\nabla \Phi)^{2} = \dot{\Phi}^{2}
\end{eqnarray}
and
\begin{eqnarray}
\Phi R &=& \omega \frac{(\nabla \Phi)^{2}}{\Phi} + 3 \nabla^{2} \Phi + 2 V \\
&=& - \left( \frac{6}{\rho^{2}} \Phi \right) (\rho \ddot{\rho} + \dot{\rho}^{2} -1),
\end{eqnarray}
and the volume factor becomes
\begin{eqnarray}
\sqrt{g} d^{4} x = 2 \pi^{2} \rho^{3} d\eta.
\end{eqnarray}

Using these equations, then, we obtain the Euclidean action as follows:
\begin{eqnarray}
S_{\,\mathrm{E}} &=& 2 \pi^{2} \int \rho^{3} d \eta \frac{1}{16\pi} \left( - \Phi R + \omega \frac{\dot{\Phi}^{2}}{\Phi} + V \right) \\
&=& \frac{\pi}{8} \int \rho^{3} d \eta \left( \frac{6}{\rho^{2}} \Phi (\rho \ddot{\rho} + \dot{\rho}^{2} -1) + \omega \frac{\dot{\Phi}^{2}}{\Phi} + V \right).
\end{eqnarray}
Upon integration by parts, we obtain
\begin{eqnarray}
S_{\,\mathrm{E}} = \frac{\pi}{8} \int d \eta \left( - 6 \dot{\Phi} \dot{\rho} \rho^{2} - 6 \Phi \rho \dot{\rho}^{2} - 6 \Phi \rho + \omega \rho^{3} \frac{\dot{\Phi}^{2}}{\Phi} + \rho^{3} V \right) + \textrm{boundary term}
\end{eqnarray}
and the boundary term is irrelevant here as we are interested in the difference between the action of an bounce and the background.
Finally, after simple calculations, we end up with
\begin{eqnarray}
S_{\,\mathrm{E}} = \frac{\pi}{4} \int d \eta \left( \rho^{3} V - 6 \rho \Phi \right).
\end{eqnarray}

This result is indeed consistent with the result of Coleman and De Luccia \cite{CDL}:
\begin{eqnarray}
S_{\,\mathrm{E}} = 4\pi^{2} \int d \eta \left( \rho^{3} V - \frac{3 \rho}{8 \pi G} \right),
\end{eqnarray}
and we obtain our result again if we change $V$ by $V/16 \pi$ and $G$ by $1/\Phi$.

Finally, if we have a solution of the Euclidean metric and field combinations, then we can approximate the probability amplitude $P$ of the Euclidean bounce by
\begin{eqnarray}
P \sim A e^{-B},
\end{eqnarray}
where
\begin{eqnarray}
B = S_{\,\mathrm{E}}(\mathrm{bounce}) - S_{\,\mathrm{E}}(\mathrm{background}).
\end{eqnarray}

\section{\label{sec:Nuc}Nucleation of vacuum bubbles in Brans-Dicke type theory}

\begin{figure}
\begin{center}
\includegraphics[scale=1.2]{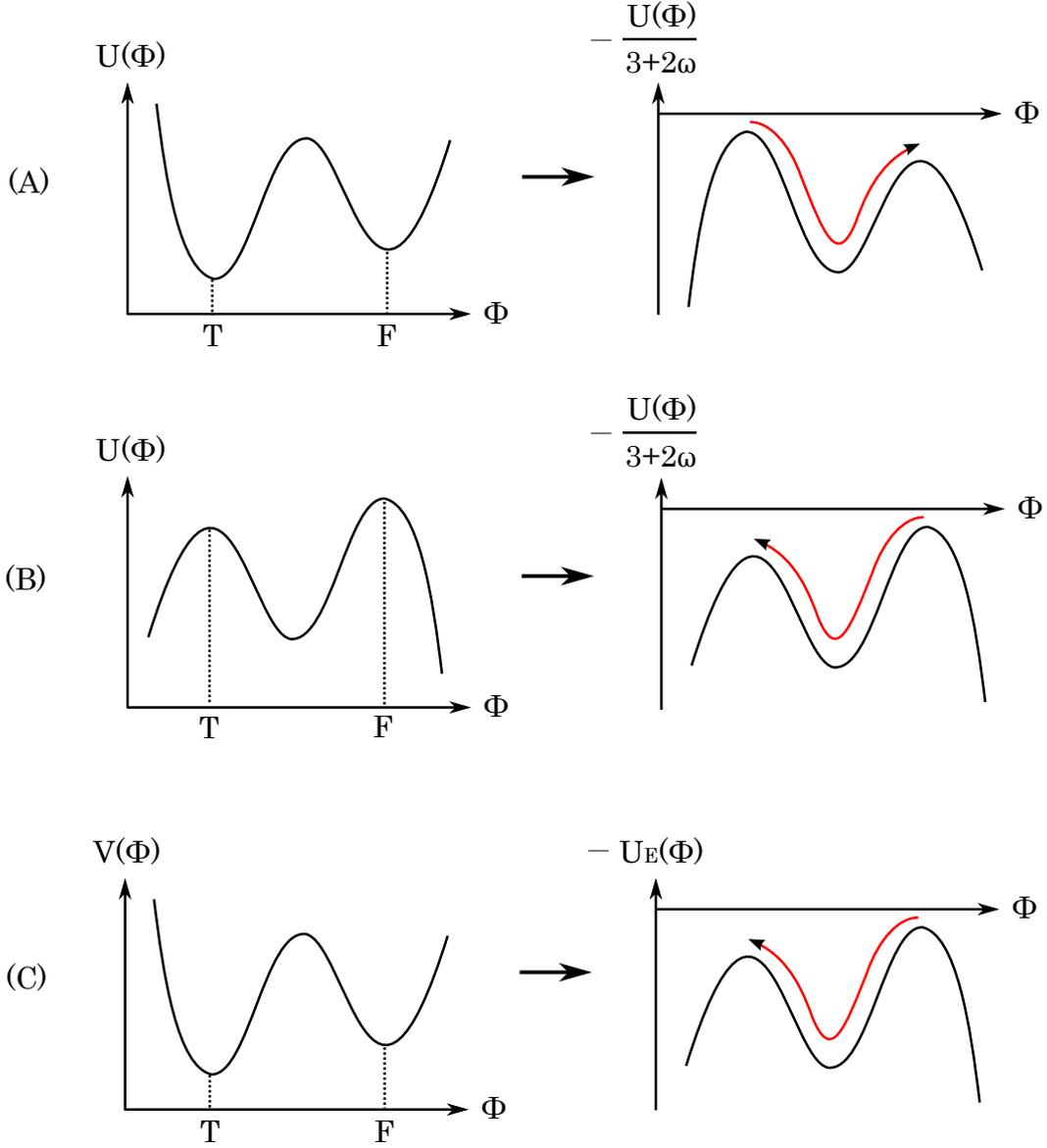}
\caption{\label{fig:bounce}Possible bounce solutions in the Brans-Dicke type theory. The Lorentzian dynamics is determined by $U(\Phi)/(3+2\omega)$, while the Euclidean dynamics is determined by $-U(\Phi)/(3+2\omega)$. $T$ and $F$ are the location of a true vacuum and a false vacuum, respectively, in the Lorentzian signature. Diagram (A) is for $\omega > -3/2$ and means a true vacuum bubble in a false vacuum background. Diagram (B) is for $\omega < -3/2$ and means a false vacuum bubble in a true vacuum background. Diagram (C) is for $\omega > -3/2$, where $V(T) < V(F)$ and $U_{E}(T) > U_{E}(F)$; hence a nucleation of a false vacuum bubble is possible.}
\end{center}
\end{figure}

In this section, we illustrate and check out possible bounce solutions of the Brans-Dicke type theory. The dynamics of the Brans-Dicke scalar field is governed by the field equation
\begin{eqnarray}
\ddot{\Phi} + 3 \frac{\dot{\rho}}{\rho} \dot{\Phi} = \frac{1}{2 \omega + 3} \frac{dU}{d\Phi} = - \frac{d}{d\Phi} \frac{-U}{2\omega+3},
\end{eqnarray}
which is, in turn, determined by $-U(\Phi)/(3+2\omega)$. The second term on the left-hand side is the damping term that eventually causes the scalar field to stop rolling for most cases. However, clearly it is the potential $V(\Phi)$ that determines the nature of the vacuum as being true or false, since the energy-momentum tensor is given by $V(\Phi)$.

In Figure~\ref{fig:bounce}, we classify possible bounce solutions. The left diagrams of Figure~\ref{fig:bounce} are typical effective potentials $U(\Phi)$ or potentials $V(\Phi)$. Let us first consider the cases $V(T) < V(F)$ and $U(T) < U(F)$, where $T$ is the field value of the true vacuum and $F$ is the field value of the false vacuum. Diagram (A) in Figure~\ref{fig:bounce} is for $\omega > -3/2$ and describes the generation of a true vacuum bubble in a false vacuum background. Diagram (B) is for $\omega < -3/2$ and describes a false vacuum bubble in a true vacuum background. However, if $V(T) < V(F)$ and $U(T) > U(F)$, then there may be a false vacuum bubble even in the $\omega > -3/2$ case. Note that the sufficient condition is not only $U(T) > U(F)$ but also $U_{E}(T) > U_{E}(F)$ (this will be confirmed in the following subsections). Diagram (C) is a situation for $\omega > -3/2$, where $V(T) < V(F)$ and $U_{E}(T) > U_{E}(F)$; hence a nucleation of a false vacuum bubble is possible.

In the following subsections, we numerically check the possibilities of the bounce solutions in detail.

\begin{table}
\begin{center}
\begin{tabular}{c|c|c|c|c|c}
\hline
Background & \;\; Bubble \;\; & \;\;\;\; $A$ \;\;\;\; & \;\;\;\; $\delta$ \;\;\;\; & \;\;\;\; $V_{0}$ \;\;\;\; & \;\;\;\; $\Phi_{\mathrm{f}}-\Phi_{\mathrm{t}}$ \;\;\;\; \\
\hline \hline
dS & dS & $10^{4}$ & $\mp$ 0.0025 & $\mathfrak{V}$ & $\mp$ 0.025 \\
\hline
dS & Flat & $10^{4}$ & $\mp$ 0.0025 & 0 & $\mp$ 0.025 \\
\hline
dS & AdS & $10^{4}$ & $\mp$ 0.0025 & $-\mathfrak{V}$ & $\mp$ 0.025 \\
\hline
Flat & AdS & $10^{4}$ & $\mp$ 0.0025 & $-2\mathfrak{V}$ & $\mp$ 0.025 \\
\hline
AdS & AdS & $10^{4}$ & $\mp$ 0.0025 & $-3\mathfrak{V}$ & $\mp$ 0.025 \\
\hline
\end{tabular}
\caption{\label{table:condition1}Potentials $V(\Phi)$ for true vacuum bubble bounces. The upper signs of $\pm$ are for $\Phi_{\mathrm{f}}<1$ (Figure~\ref{fig:lessTV}) and the lower signs of $\pm$ are for $\Phi_{\mathrm{f}}>1$ (Figure~\ref{fig:moreTV}). Here, $\mathfrak{V}=3.092\times10^{-5}$ for $\Phi_{\mathrm{f}}<1$ and $\mathfrak{V}=3.418\times10^{-5}$ for $\Phi_{\mathrm{f}}>1$.}
\end{center}
\end{table}
\begin{table}
\begin{center}
\begin{tabular}{c|c|c|c|c|c}
\hline
Background & \;\; Bubble \;\; & \;\;\;\; $A$ \;\;\;\; & \;\;\;\; $\delta$ \;\;\;\; & \;\;\;\; $V_{0}$ \;\;\;\; & \;\;\;\; $\Phi_{\mathrm{f}}-\Phi_{\mathrm{t}}$ \;\;\;\; \\
\hline \hline
dS & dS & $-10^{4}$ & $\pm$ 0.0025 & $\mathfrak{V}$ & $\mp$ 0.025 \\
\hline
Flat & dS & $-10^{4}$ & $\pm$ 0.0025 & 0 & $\mp$ 0.025 \\
\hline
AdS & dS & $-10^{4}$ & $\pm$ 0.0025 & $-\mathfrak{V}$ & $\mp$ 0.025 \\
\hline
AdS & Flat & $-10^{4}$ & $\pm$ 0.0025 & $-2\mathfrak{V}$ & $\mp$ 0.025 \\
\hline
AdS & AdS & $-10^{4}$ & $\pm$ 0.0025 & $-3\mathfrak{V}$ & $\mp$ 0.025 \\
\hline
\end{tabular}
\caption{\label{table:condition2}Potentials $V(\Phi)$ for false vacuum bubble bounces. The upper signs of $\pm$ and $\mp$ are for $\Phi_{\mathrm{f}}<1$ (Figure~\ref{fig:lessFV}) and the lower signs of $\pm$ and $\mp$ are for $\Phi_{\mathrm{f}}>1$ (Figure~\ref{fig:moreFV}). Here, $\mathfrak{V}=3.336\times10^{-5}$ for $\Phi_{\mathrm{f}}<1$ and $\mathfrak{V}=3.174\times10^{-5}$ for $\Phi_{\mathrm{f}}>1$.}
\end{center}
\end{table}

\begin{figure}
\begin{center}
\includegraphics[scale=0.6]{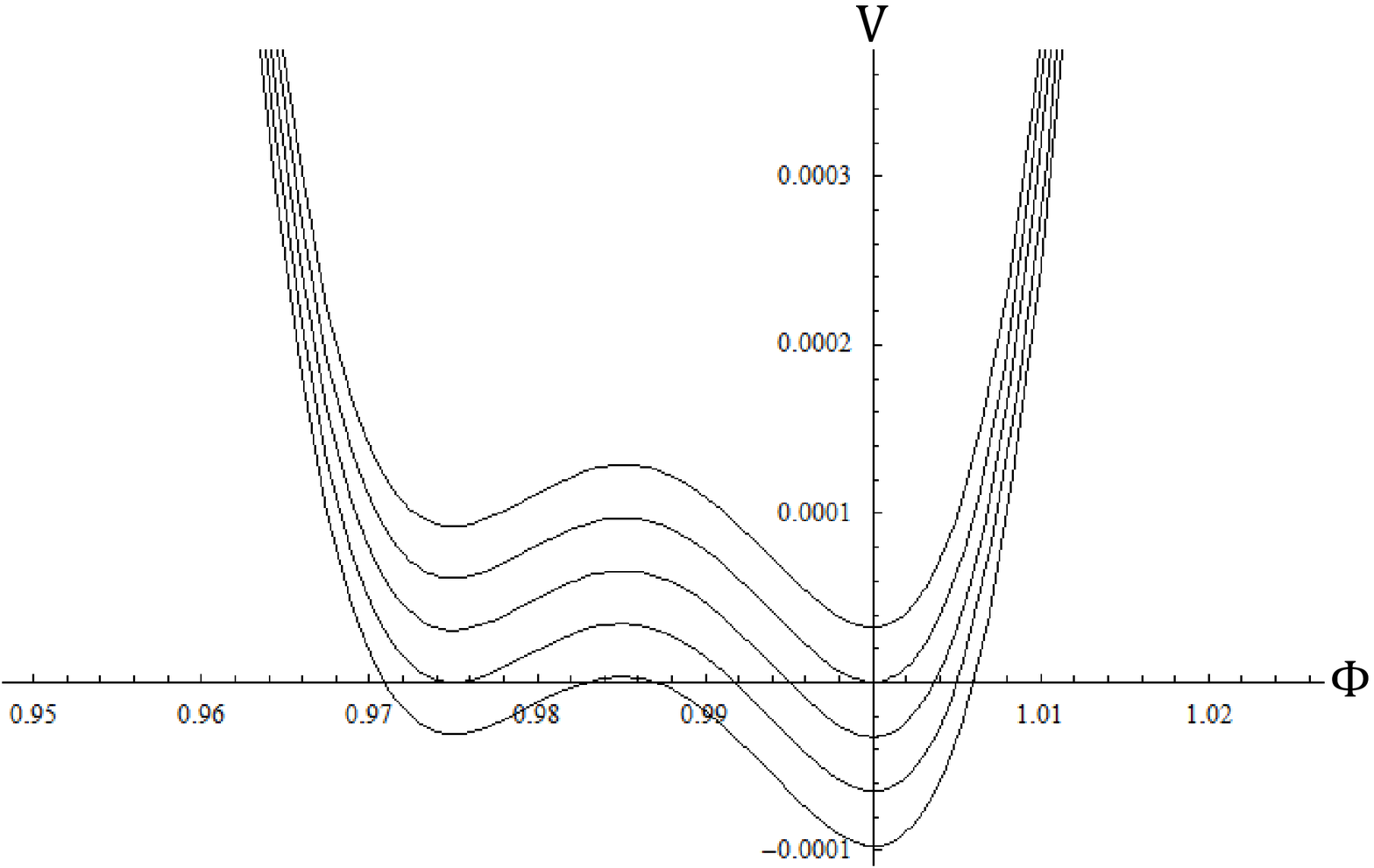}
\caption{\label{fig:lessTV}Potentials $V(\Phi)$ for $A=10^{4}$, $\delta=-0.0025$, $\Phi_{\mathrm{f}}-\Phi_{\mathrm{t}} = - 0.025$, and hence for $\Phi_{\mathrm{f}}<1$. We choose $V_{0}$ as in Table~\ref{table:condition1} to vary the true vacuum energy. From top to bottom, each potential describes a de Sitter background and a de Sitter bubble, a de Sitter background and a flat bubble, a de Sitter background and an anti-de Sitter bubble, a flat background and an anti-de Sitter bubble, and an anti-de Sitter background and an anti-de Sitter bubble.}
\end{center}
\end{figure}
\begin{figure}
\begin{center}
\includegraphics[scale=0.6]{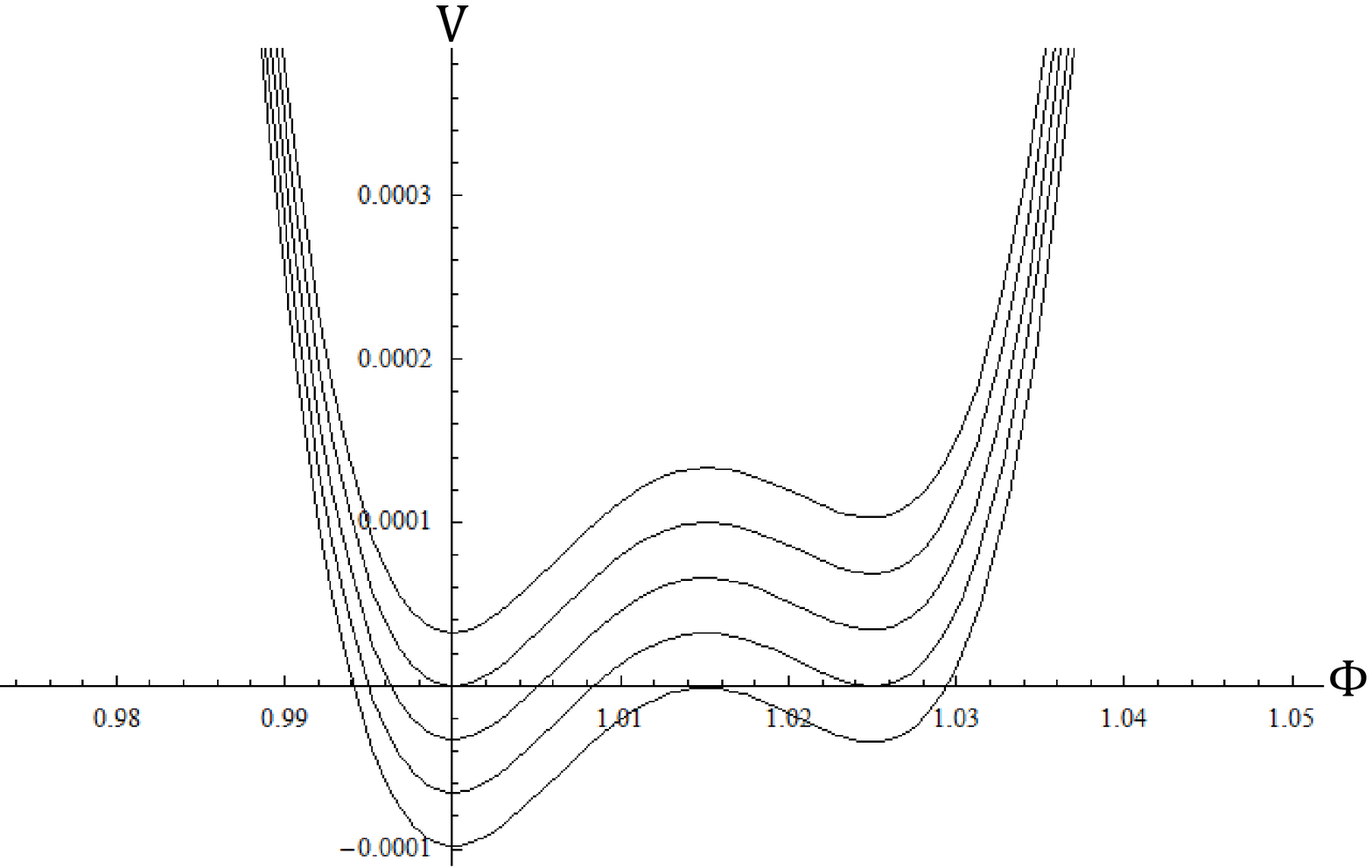}
\caption{\label{fig:moreTV}Potentials $V(\Phi)$ for $A=10^{4}$, $\delta=0.0025$, $\Phi_{\mathrm{f}}-\Phi_{\mathrm{t}} = 0.025$, and hence for $\Phi_{\mathrm{f}}>1$. We choose $V_{0}$ as in Table~\ref{table:condition1} to vary the true vacuum energy.}
\end{center}
\end{figure}

\begin{figure}
\begin{center}
\includegraphics[scale=0.8]{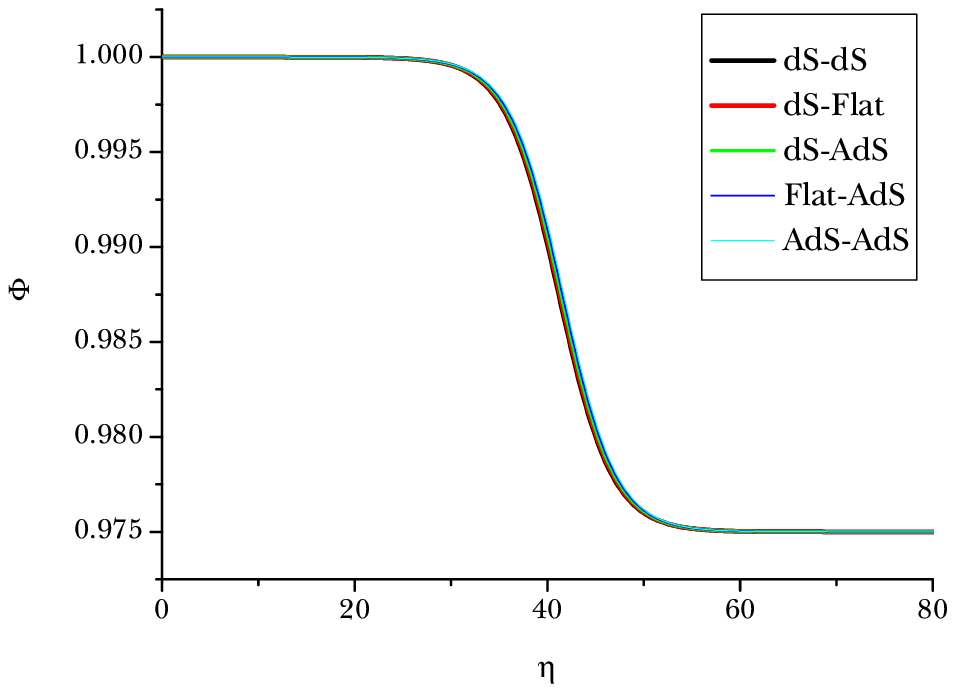}
\includegraphics[scale=0.8]{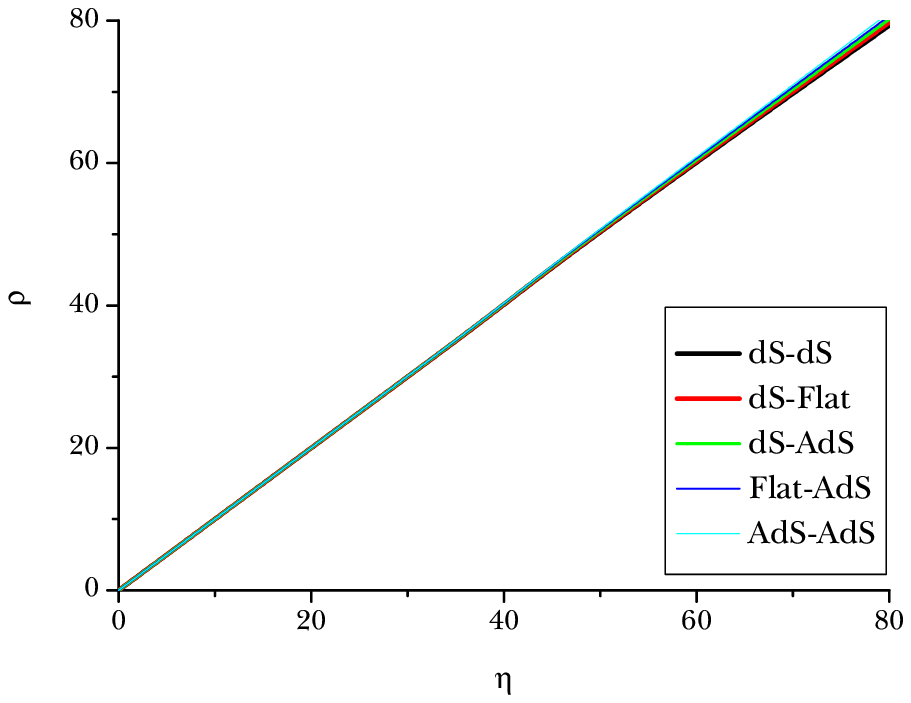}
\includegraphics[scale=0.8]{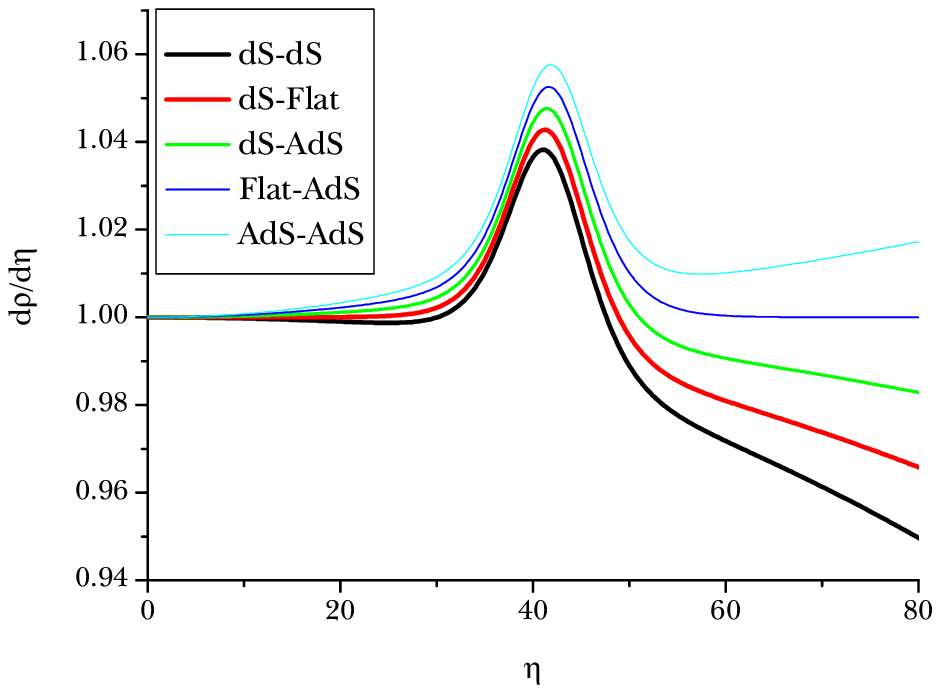}
\caption{\label{fig:bounce_lessTV}For $\omega > -3/2$, we illustrate bounce solutions for true vacuum bubbles in false vacuum backgrounds by the potentials in Figure~\ref{fig:lessTV} ($\Phi_{\mathrm{f}}<1$). Initial conditions are in Table~\ref{table:false} ($\delta < 0$ and $\Phi_{\mathrm{f}}-\Phi_{\mathrm{t}} < 0$).  We plot $\Phi$, $\rho$, and $\dot{\rho}$ as functions of $\eta$. Each caption for each curve describes a background and a bubble (e.g., $\mathrm{dS}-\mathrm{dS}$ describes a de Sitter background and a de Sitter bubble).}
\end{center}
\end{figure}
\begin{figure}
\begin{center}
\includegraphics[scale=0.8]{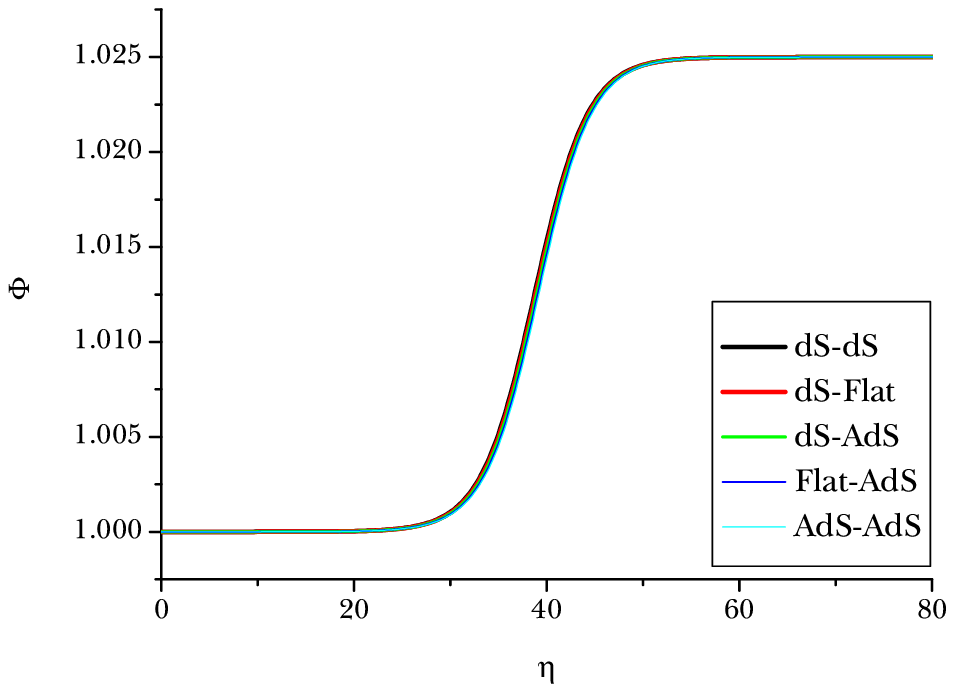}
\includegraphics[scale=0.8]{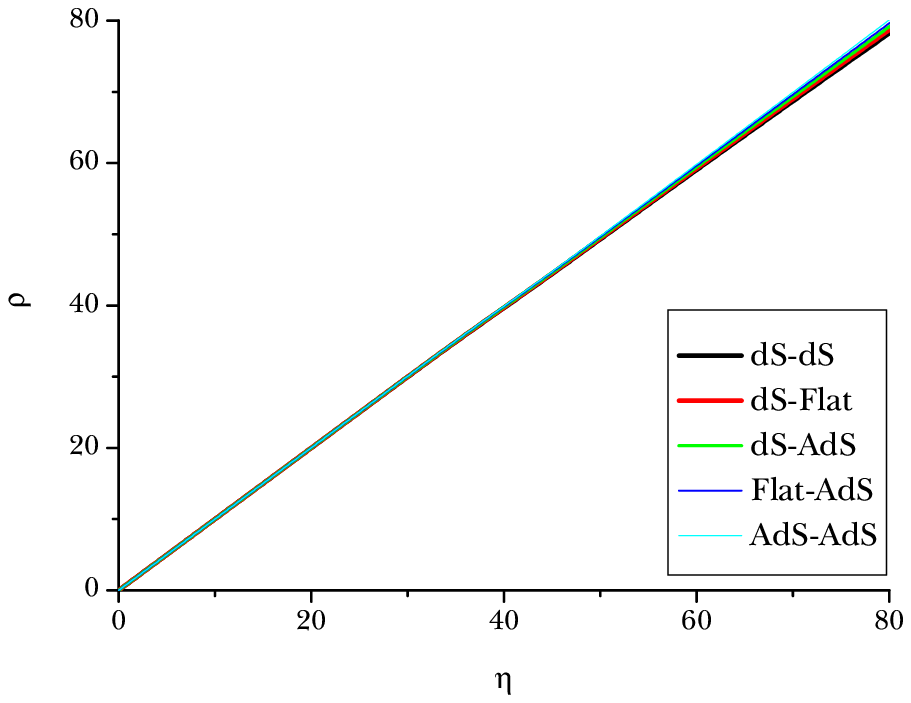}
\includegraphics[scale=0.8]{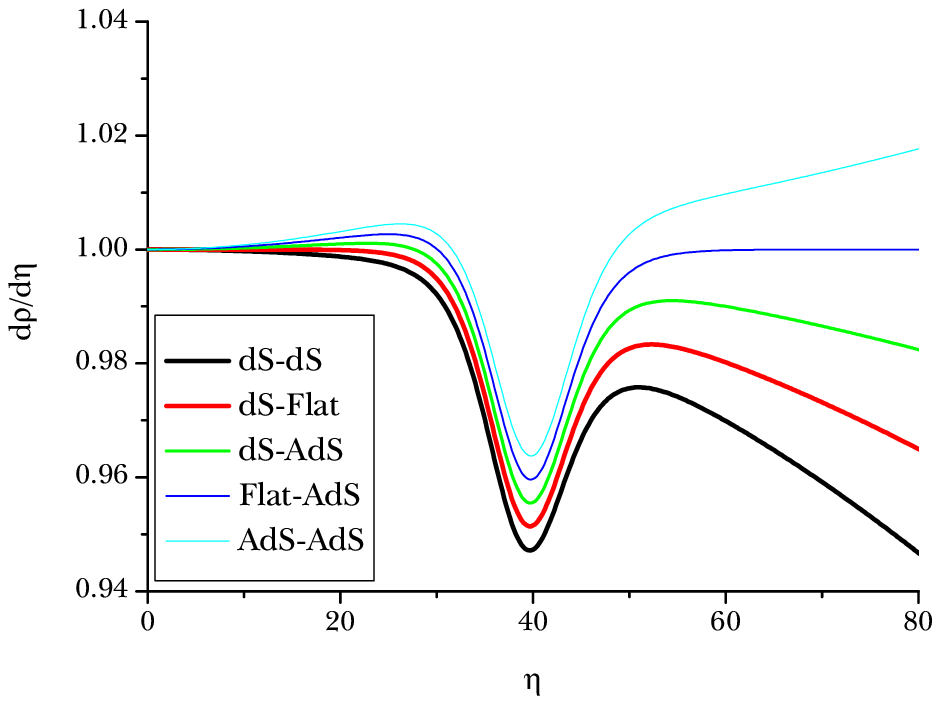}
\caption{\label{fig:bounce_moreTV}For $\omega > -3/2$, we illustrate bounce solutions for true vacuum bubbles in false vacuum backgrounds by the potentials in Figure~\ref{fig:lessTV} ($\Phi_{\mathrm{f}}<1$). Initial conditions are in Table~\ref{table:false} ($\delta > 0$ and $\Phi_{\mathrm{f}}-\Phi_{\mathrm{t}} > 0$).  We plot $\Phi$, $\rho$, and $\dot{\rho}$ as functions of $\eta$. Each caption for each curve describes a background and a bubble (e.g., $\mathrm{dS}-\mathrm{dS}$ describes a de Sitter background and a de Sitter bubble).}
\end{center}
\end{figure}

\subsection{$\omega > -3/2$: true vacuum bubbles in false vacuum backgrounds}

First, we consider true vacuum bubbles in false vacuum backgrounds. There are five possibilities: a de Sitter (dS) bubble in a de Sitter background, a flat bubble in a de Sitter background, an anti-de Sitter bubble in a de Sitter background, an anti-de Sitter bubble in a flat background, and an-anti de Sitter bubble in an-anti de Sitter background. Also, there are two possibilities for $\Phi_{\mathrm{f}}$, depending on whether it is more or less than $1$. Therefore, to study these possibilities, we considered ten potentials, as illustrated in Table~\ref{table:condition1}. Here, we used $\omega=10$, and hence it is greater than $-3/2$. Figures~\ref{fig:lessTV} and \ref{fig:moreTV} show the potentials we used.

\subsubsection{$\Phi_{\mathrm{f}}<1$}

In Figure~\ref{fig:bounce_lessTV}, we denote bounce solutions for true vacuum bubbles in false vacuum backgrounds by potentials in Figure~\ref{fig:lessTV} ($\Phi_{\mathrm{f}}<1$). We plot $\Phi$, $\rho$, and $\dot{\rho}$ as functions of $\eta$.

\subsubsection{$\Phi_{\mathrm{f}}>1$}

In Figure~\ref{fig:bounce_moreTV}, we denote bounce solutions for true vacuum bubbles in false vacuum backgrounds by potentials in Figure~\ref{fig:moreTV} ($\Phi_{\mathrm{f}}>1$). We plot $\Phi$, $\rho$, and $\dot{\rho}$ as functions of $\eta$.

Note that $\rho$ is a $\sin$ function for a de Sitter space, proportional to $\eta$ for a flat space, and a $\sinh$ function for an anti-de Sitter space. Therefore, $\dot{\rho}$ is a $\cos$ function for a de Sitter space, $1$ for a flat space, and a $\cosh$ function for an anti-de Sitter space. In our results, $\rho$ is too close to compare, but $\dot{\rho}$ can be distinguished. Such behaviors ($\cos$, $1$, $\cosh$, etc.) consistently hold for inside and outside of the transition region.

\begin{figure}
\begin{center}
\includegraphics[scale=0.6]{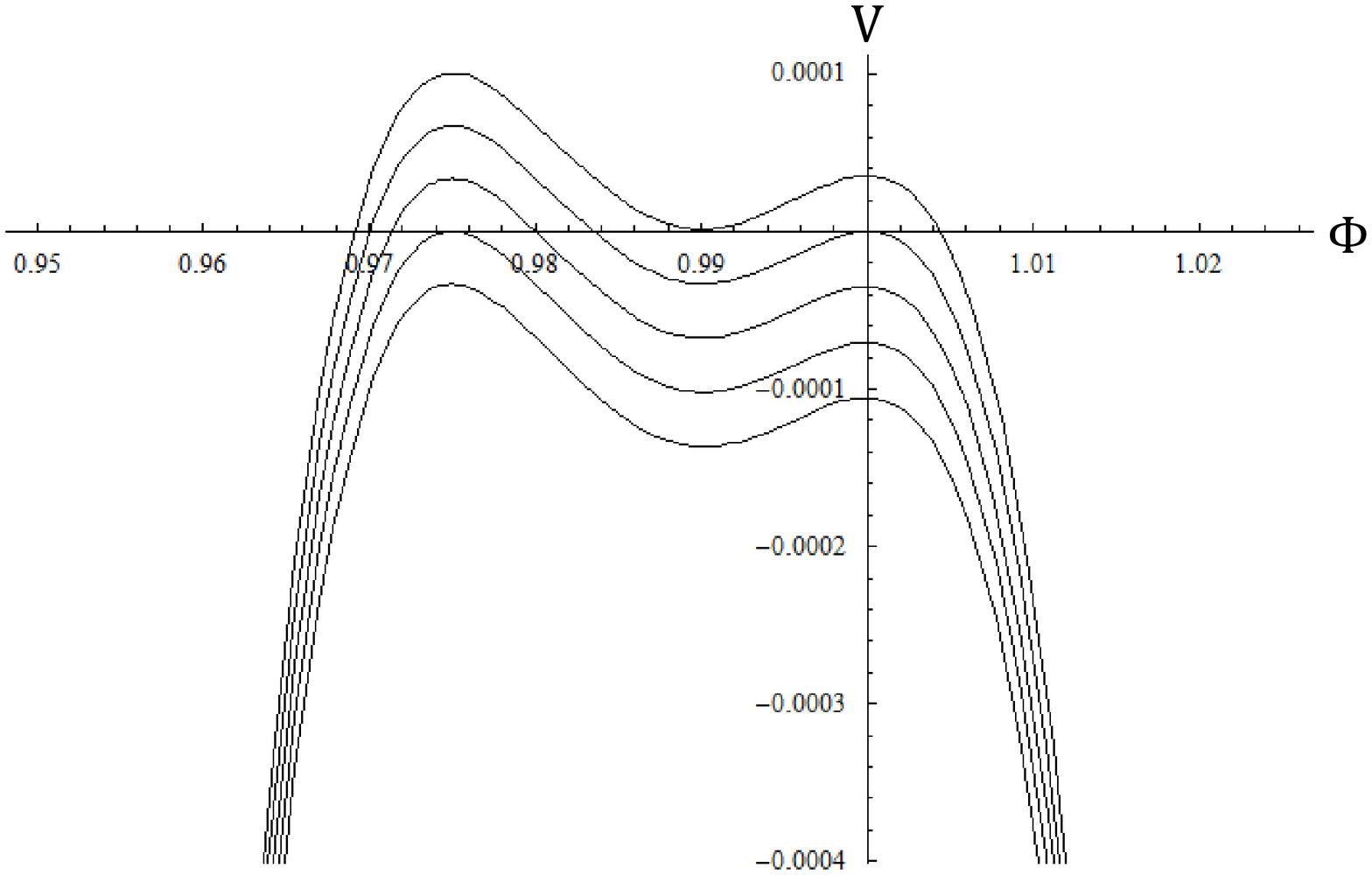}
\caption{\label{fig:lessFV}Potentials $V(\Phi)$ for $A=-10^{4}$, $\delta=0.0025$, $\Phi_{\mathrm{f}}-\Phi_{\mathrm{t}} = - 0.025$, and hence for $\Phi_{\mathrm{f}}<1$. We choose $V_{0}$ as in Table~\ref{table:condition2} to vary the true vacuum energy. From top to bottom, each potential means a de Sitter background and a de Sitter bubble, a flat background and a de Sitter bubble, an anti-de Sitter background and a de Sitter bubble, an anti-de Sitter background and a flat bubble, and an anti-de Sitter background and an anti-de Sitter bubble.}
\end{center}
\end{figure}
\begin{figure}
\begin{center}
\includegraphics[scale=0.6]{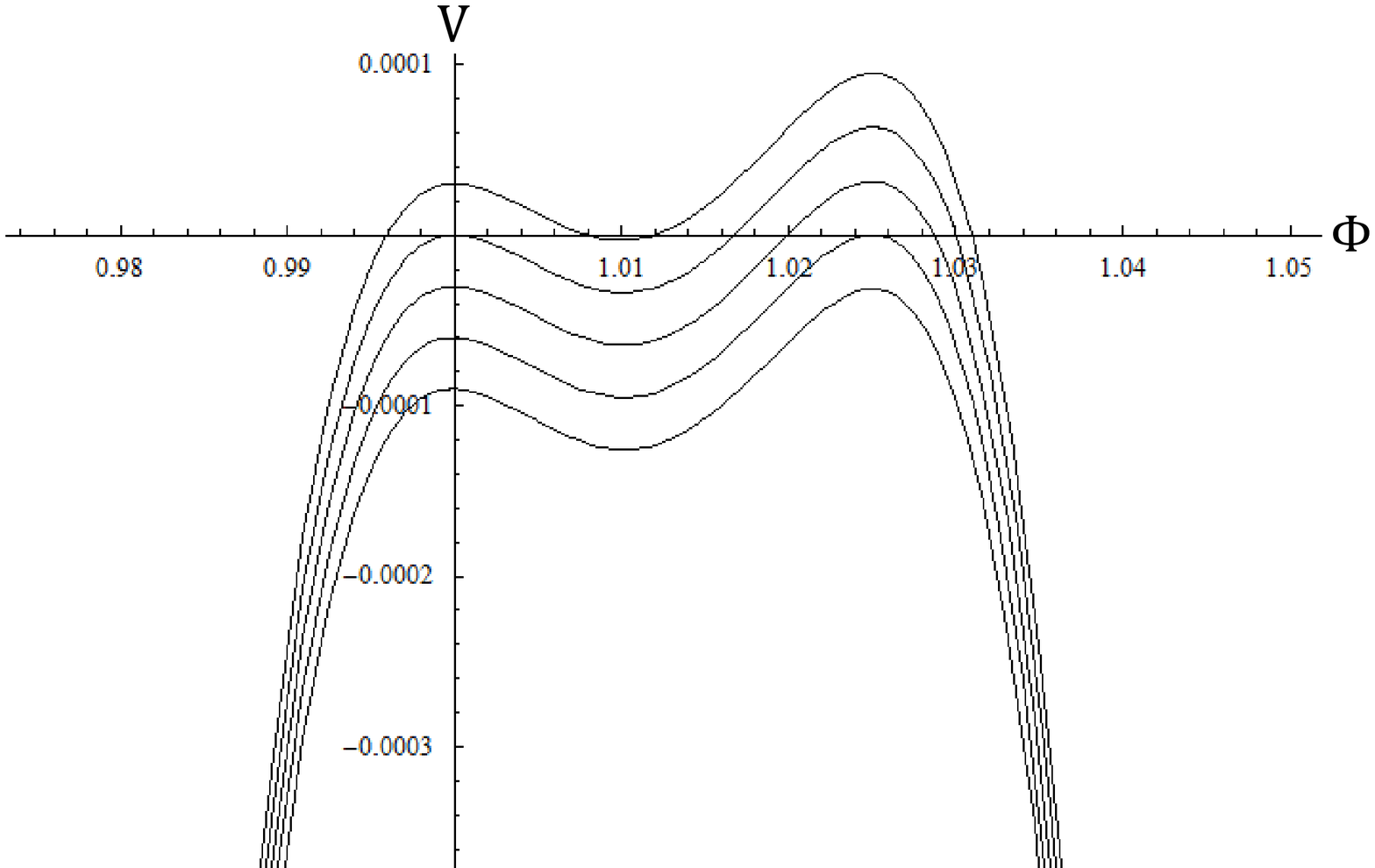}
\caption{\label{fig:moreFV}Potentials $V(\Phi)$ for $A=-10^{4}$, $\delta=-0.0025$, $\Phi_{\mathrm{f}}-\Phi_{\mathrm{t}} = 0.025$, and hence for $\Phi_{\mathrm{f}}>1$. We choose $V_{0}$ as in Table~\ref{table:condition2} to vary the true vacuum energy.}
\end{center}
\end{figure}

\begin{figure}
\begin{center}
\includegraphics[scale=0.8]{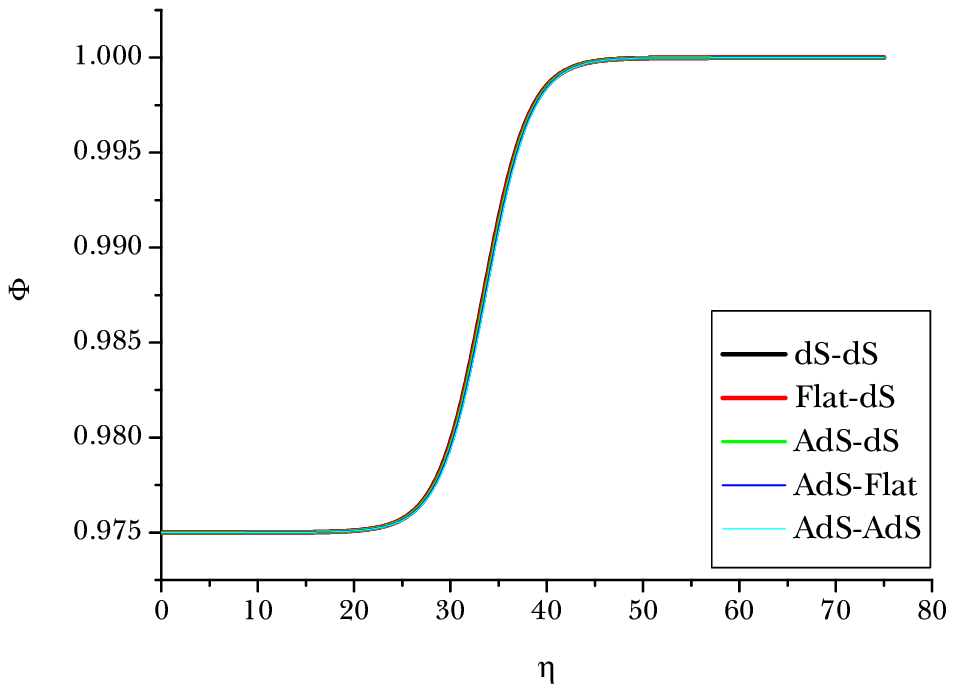}
\includegraphics[scale=0.8]{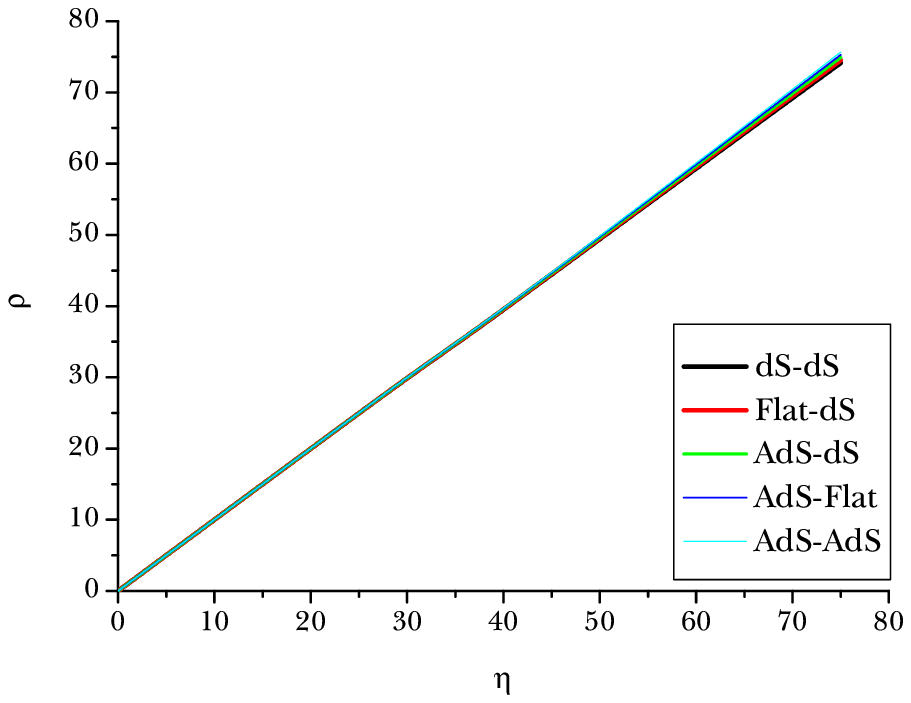}
\includegraphics[scale=0.8]{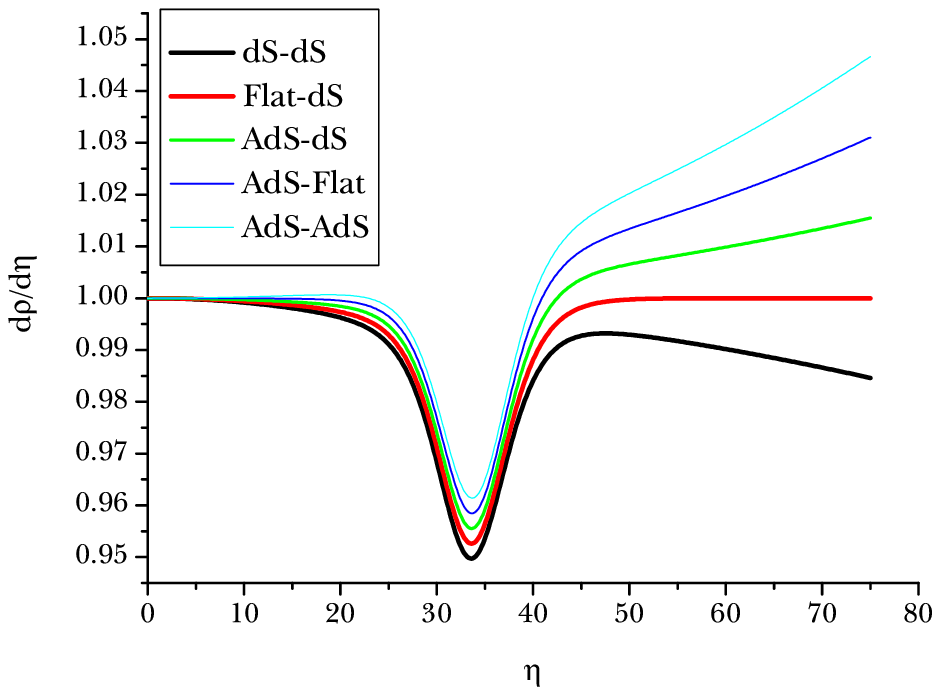}
\caption{\label{fig:bounce_lessFV}For $\omega < -3/2$, we illustrate bounce solutions for false vacuum bubbles in true vacuum backgrounds by the potentials in Figure~\ref{fig:lessFV} ($\Phi_{\mathrm{f}}<1$). Initial conditions are in Table~\ref{table:false} ($\delta>0$ and $\Phi_{\mathrm{f}}-\Phi_{\mathrm{t}} < 0$).  We plot $\Phi$, $\rho$, and $\dot{\rho}$ as functions of $\eta$. Each caption for each curve means a background and a bubble (e.g., $\mathrm{dS}-\mathrm{dS}$ means a de Sitter background and a de Sitter bubble).}
\end{center}
\end{figure}
\begin{figure}
\begin{center}
\includegraphics[scale=0.8]{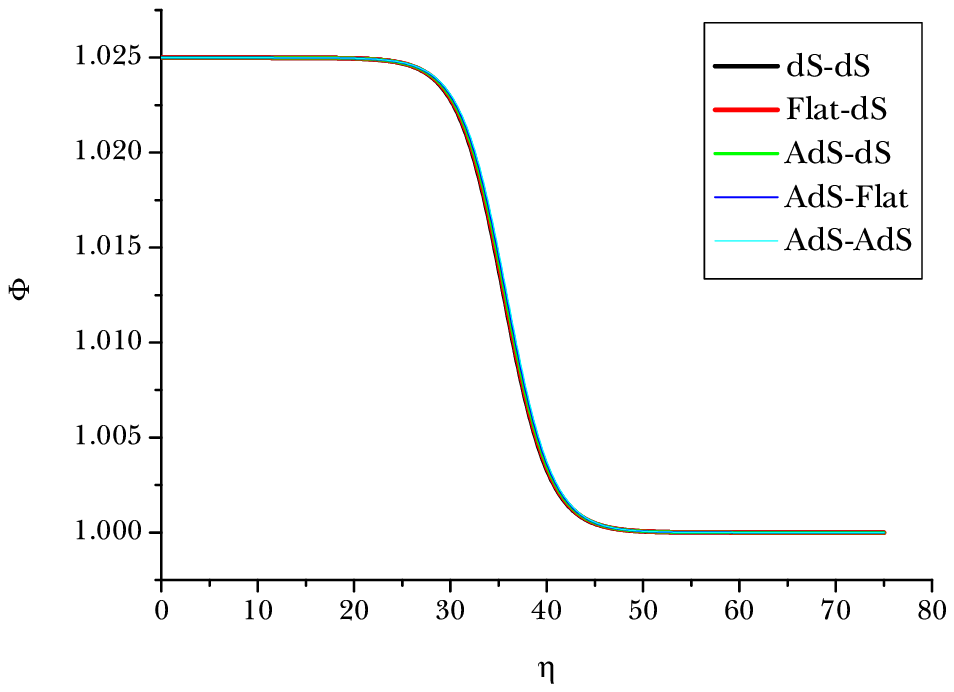}
\includegraphics[scale=0.8]{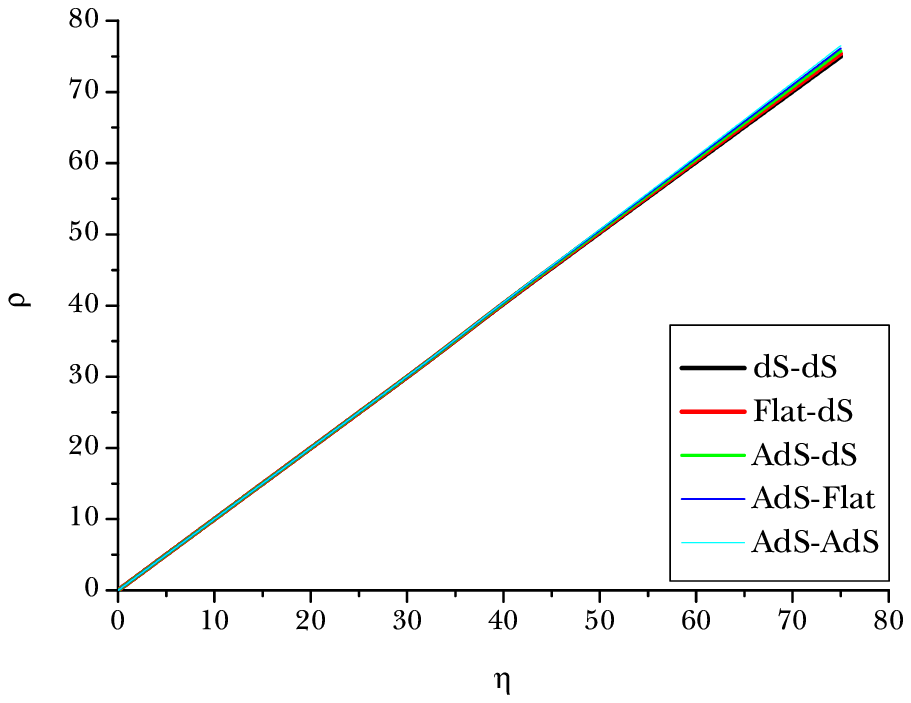}
\includegraphics[scale=0.8]{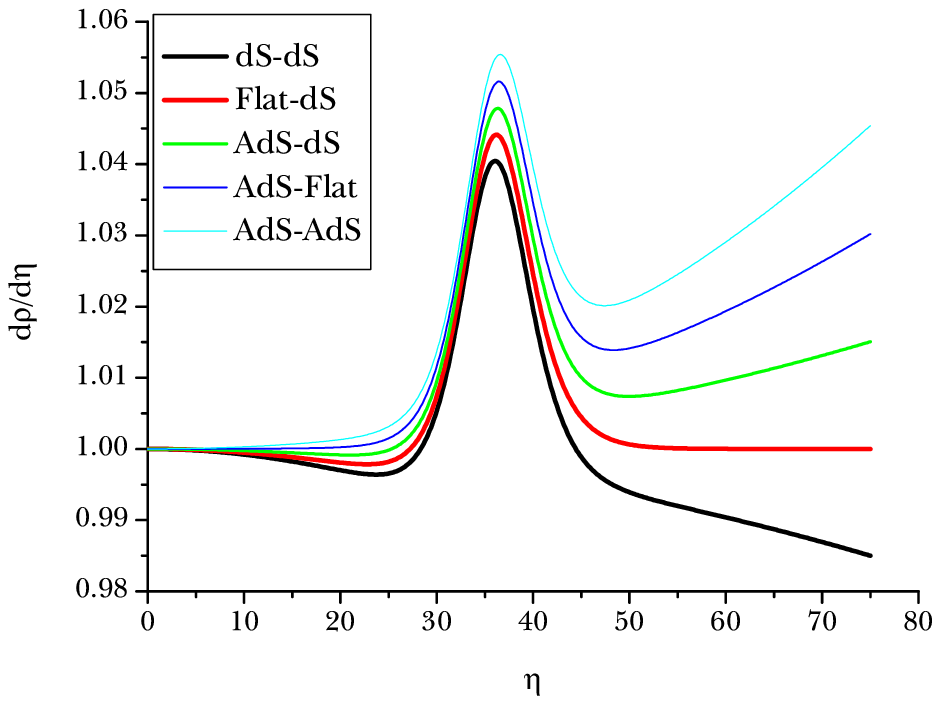}
\caption{\label{fig:bounce_moreFV}For $\omega < -3/2$, we illustrate bounce solutions for false vacuum bubbles in true vacuum backgrounds by the potentials in Figure~\ref{fig:moreFV} ($\Phi_{\mathrm{f}}>1$). Initial conditions are in Table~\ref{table:false} ($\delta<0$ and $\Phi_{\mathrm{f}}-\Phi_{\mathrm{t}} > 0$).  We plot $\Phi$, $\rho$, and $\dot{\rho}$ as functions of $\eta$. Each caption for each curve means a background and a bubble (e.g., $\mathrm{dS}-\mathrm{dS}$ means a de Sitter background and a de Sitter bubble).}
\end{center}
\end{figure}

\subsection{$\omega < -3/2$: false vacuum bubbles in true vacuum backgrounds}

Second, we consider false vacuum bubbles in true vacuum backgrounds. There are five possibilities: a de Sitter bubble in a de Sitter background, a de Sitter bubble in a flat background, a de Sitter bubble in an-anti de Sitter background, a flat bubble in an anti-de Sitter background, and an anti-de Sitter bubble in an anti-de Sitter background. Also, there are two possibilities for $\Phi_{\mathrm{f}}$, depending on whether it is more or less than $1$. Therefore, to study these possibilities, we considered ten potentials, as illustrated in Table~\ref{table:condition2}. Here, we used $\omega=-2$, and hence it is less than $-3/2$. Figures~\ref{fig:lessFV} and \ref{fig:moreFV} show the potentials we used.

It should be noted that two peaks in potentials in Figures~\ref{fig:lessFV} and \ref{fig:moreFV} are \textit{stable} vacua for $\omega < -3/2$. This is obvious since the dynamics of the field is determined by the field equation in the Lorentzian signatures
\begin{eqnarray}
\ddot{\Phi} + 3 \frac{\dot{\rho}}{\rho} \dot{\Phi} = - \frac{1}{2 \omega + 3} \frac{dU}{d\Phi} = \frac{1}{| 2 \omega + 3 |} \frac{dU}{d\Phi},
\end{eqnarray}
and hence it is determined by $-U(\Phi)/| 2 \omega + 3 |$.

\subsubsection{$\Phi_{\mathrm{f}}<1$}

In Figure~\ref{fig:bounce_lessFV}, we denote bounce solutions for false vacuum bubbles in true vacuum backgrounds by potentials in Figure~\ref{fig:lessFV} ($\Phi_{\mathrm{f}}<1$). We plot $\Phi$, $\rho$, and $\dot{\rho}$ as functions of $\eta$.

\subsubsection{$\Phi_{\mathrm{f}}>1$}

In Figure~\ref{fig:bounce_moreFV}, we denote bounce solutions for false vacuum bubbles in true vacuum backgrounds by potentials in Figure~\ref{fig:moreFV} ($\Phi_{\mathrm{f}}>1$). We plot $\Phi$, $\rho$, and $\dot{\rho}$ as functions of $\eta$.

We also note that behaviors of $\dot{\rho}$ ($\cos$, $1$, $\cosh$, etc.) are consistent for the inside and the outside of the transition region.

\subsection{\label{sec:via}False vacuum bubble nucleation via effective potentials}

Although $V(\Phi_{\mathrm{t}}) < V(\Phi_{\mathrm{f}})$, if the vacuum energy of $\Phi_{\mathrm{f}}$ in the Einstein frame is smaller than that of $\Phi_{\mathrm{t}}$, i.e., $U_{E}(\Phi_{\mathrm{t}}) > U_{E}(\Phi_{\mathrm{f}})$, where $U_{E}$ is the potential in the Einstein frame, then, interestingly, it may be possible to obtain a false vacuum bubble in the Jordan frame, even if $\omega > -3/2$. Note that the dynamics of the Brans-Dicke field is determined by the effective potential $U$. Hence, we also should check whether $U(\Phi_{\mathrm{t}}) > U(\Phi_{\mathrm{f}})$.

Note that such conditions can be represented as
\begin{eqnarray}
V(\Phi_{\mathrm{f}})-V(\Phi_{\mathrm{t}}) = \Phi_{\mathrm{f}}^{2} \left( \int_{1}^{\Phi_{\mathrm{f}}} \frac{F(\bar{\Phi})}{\bar{\Phi}^{3}} d \bar{\Phi} + V_{0} \right) - V_{0} > 0
\end{eqnarray}
and
\begin{eqnarray}
U_{E}(\Phi_{\mathrm{f}})-U_{E}(\Phi_{\mathrm{t}}) = \int_{1}^{\Phi_{\mathrm{f}}} \frac{F(\bar{\Phi})}{\bar{\Phi}^{3}} d \bar{\Phi} \equiv \Delta E < 0.
\end{eqnarray}
Therefore, we require $V_{0} > \Phi^{2}_{\mathrm{f}} |\Delta E|/ (\Phi^{2}_{\mathrm{f}}-1)$ and we conclude that such false vacuum bubbles can form only in a de Sitter space background ($V_{0}>0$) if $\Phi_{\mathrm{f}} > 1$. (In the next section, we shall discuss that a false vacuum bubble can expand in the Lorentzian signatures only if $\Phi_{\mathrm{f}} > 1$. Hence, we only consider this case.)

\begin{figure}
\begin{center}
\includegraphics[scale=0.6]{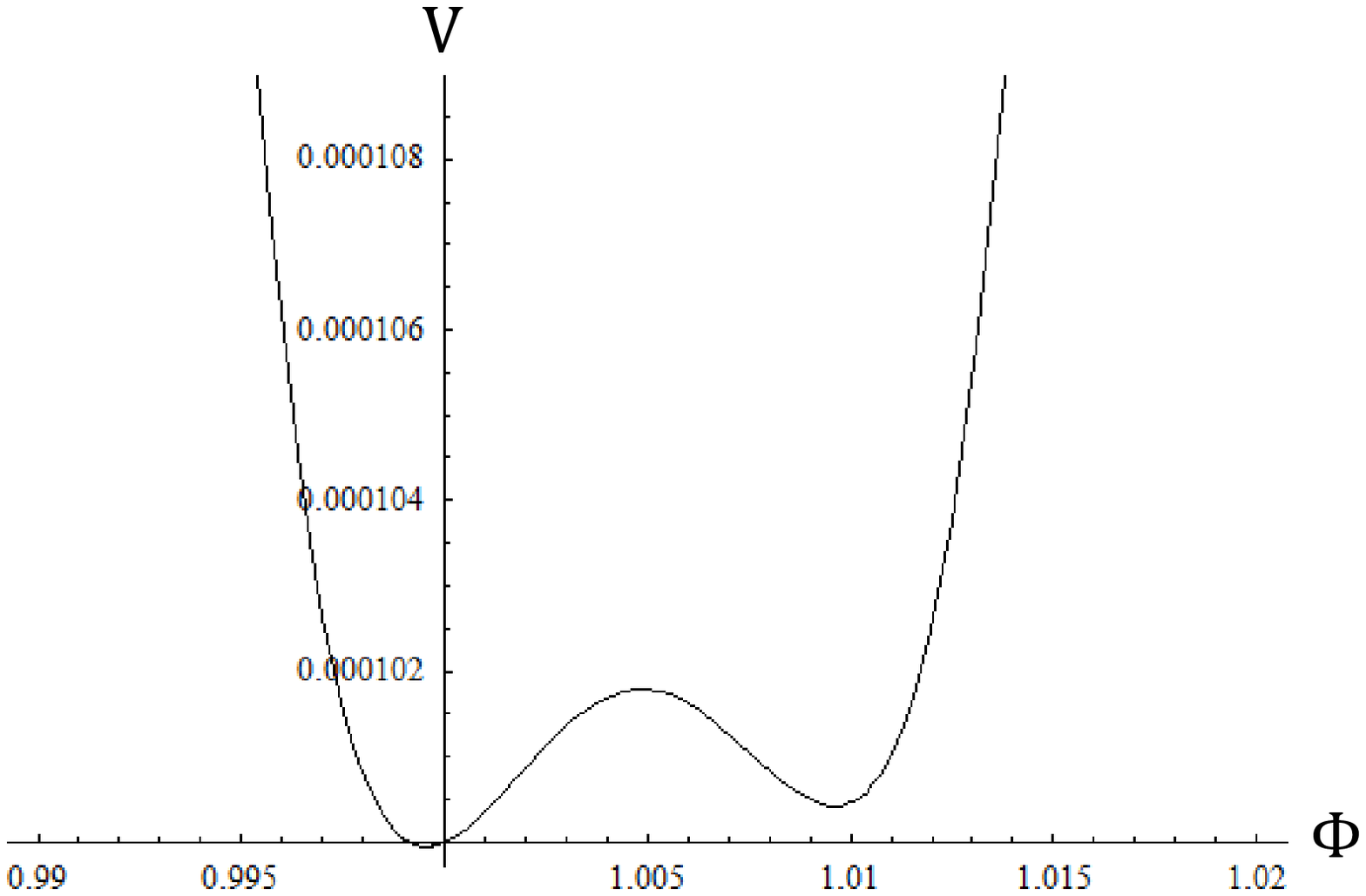}
\includegraphics[scale=0.6]{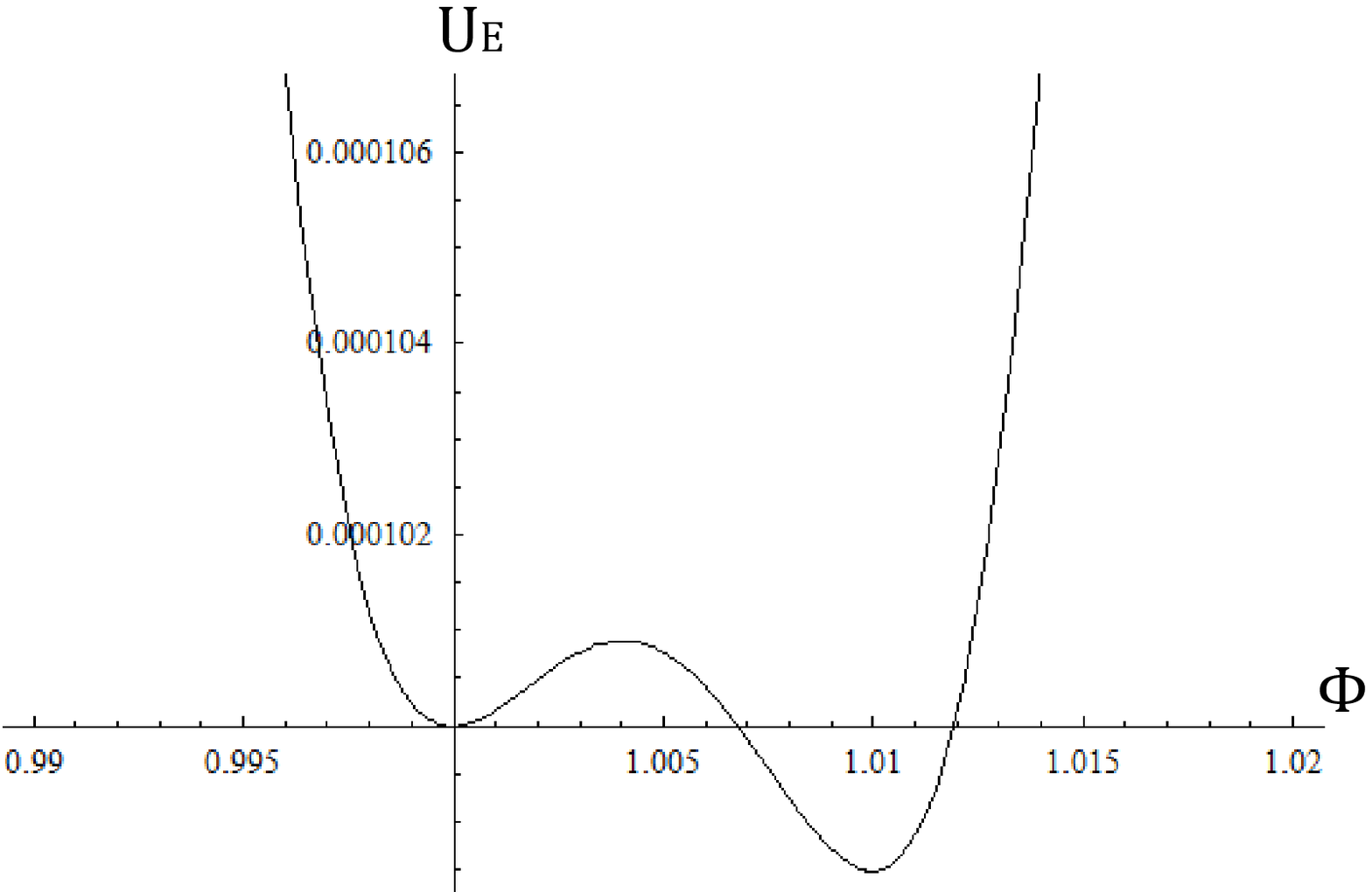}
\includegraphics[scale=0.6]{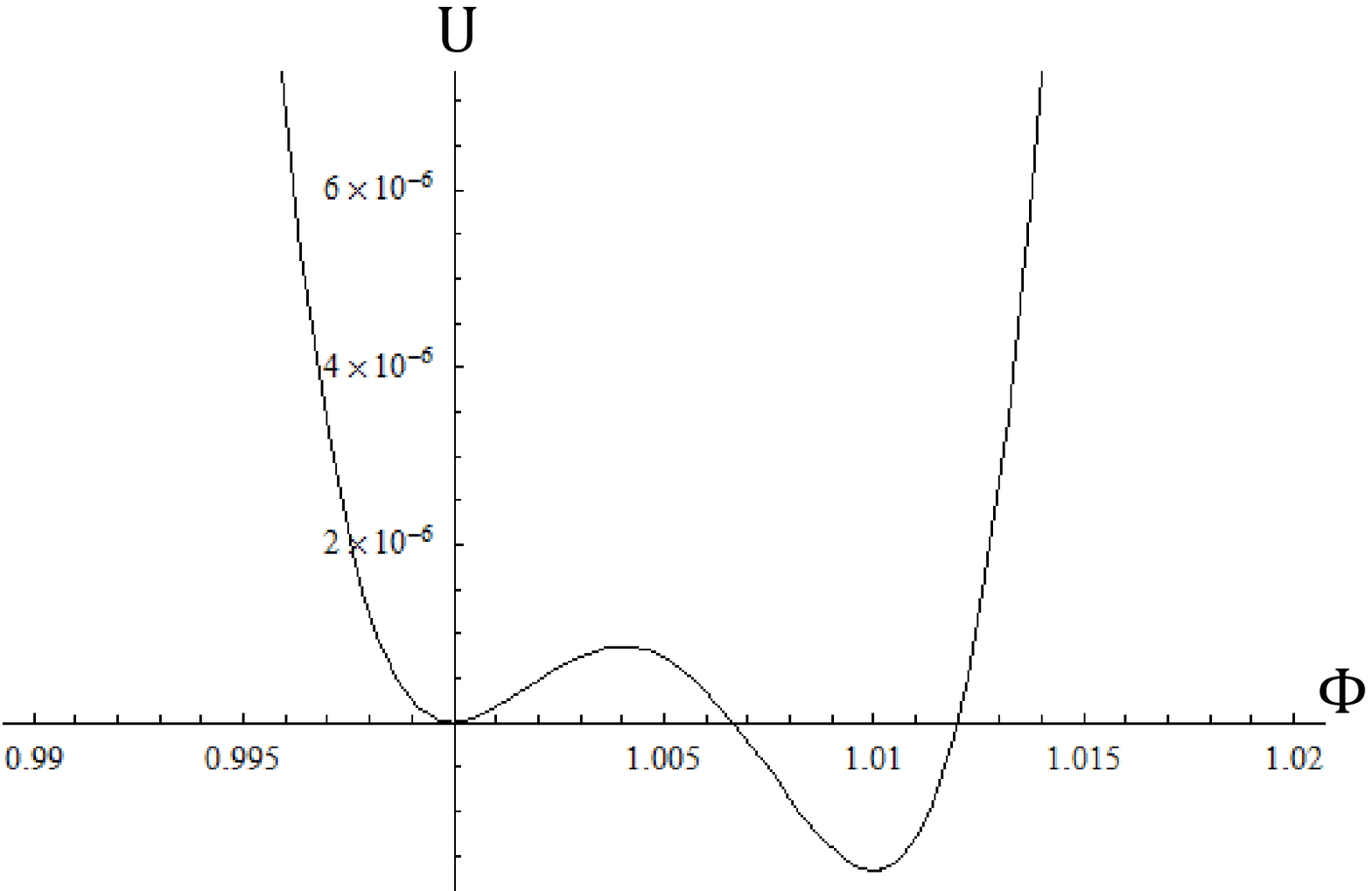}
\caption{\label{fig:type3_potential}For $\omega = 10$, $A=10^{4}$, $\Phi_{\mathrm{f}}-\Phi_{\mathrm{t}} = 0.01$, $\delta = -0.001$, and $V_{0}=0.0001$, we plot potentials $V$, $U_{E}$, and $U$.}
\end{center}
\end{figure}

If we choose parameters as $\omega = 10$, $A=10^{4}$, $\Phi_{\mathrm{f}}-\Phi_{\mathrm{t}} = 0.01$, $\delta = -0.001$, and $V_{0}=0.0001$, we obtain potentials for such conditions (Figure~\ref{fig:type3_potential}). Here, we plot $V$, $U_{E}$, and $U$. We find that $V(\Phi_{\mathrm{t}}) < V(\Phi_{\mathrm{f}})$, and hence $\Phi_{\mathrm{f}}$ is in a false vacuum in the Jordan frame, but $U_{E}(\Phi_{\mathrm{t}}) > U_{E}(\Phi_{\mathrm{f}})$ and $U(\Phi_{\mathrm{t}}) > U(\Phi_{\mathrm{f}})$. Therefore, $\Phi_{\mathrm{f}}$ is in a true vacuum in the Einstein frame. If a true vacuum bubble can form in the Einstein frame \cite{CDL}, it will correspond to a false vacuum bubble in the Jordan frame.

\begin{figure}
\begin{center}
\includegraphics[scale=0.8]{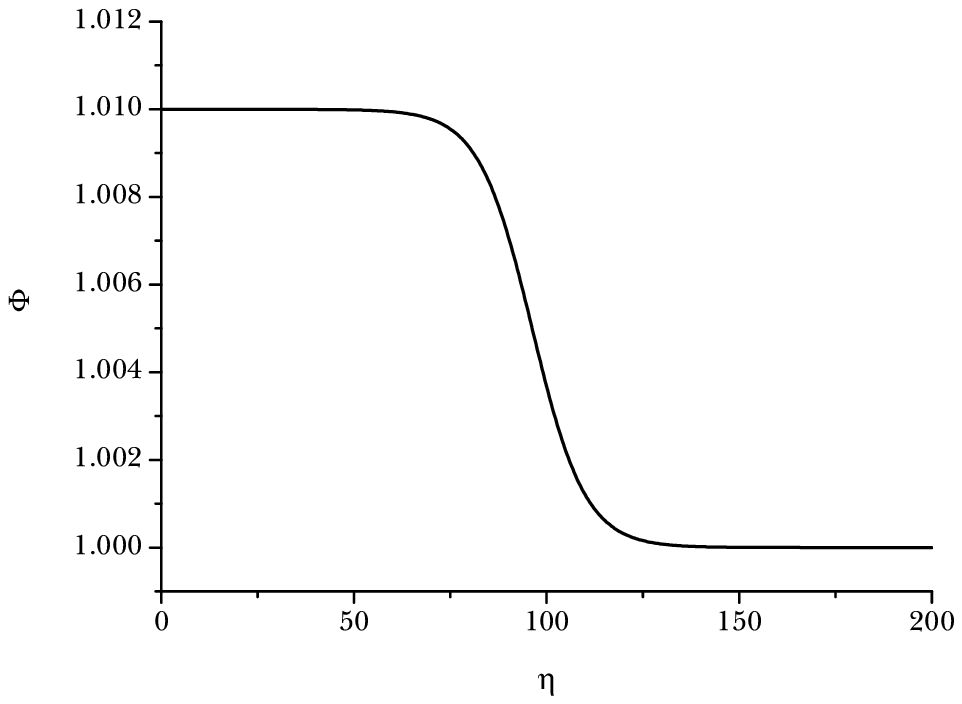}
\includegraphics[scale=0.8]{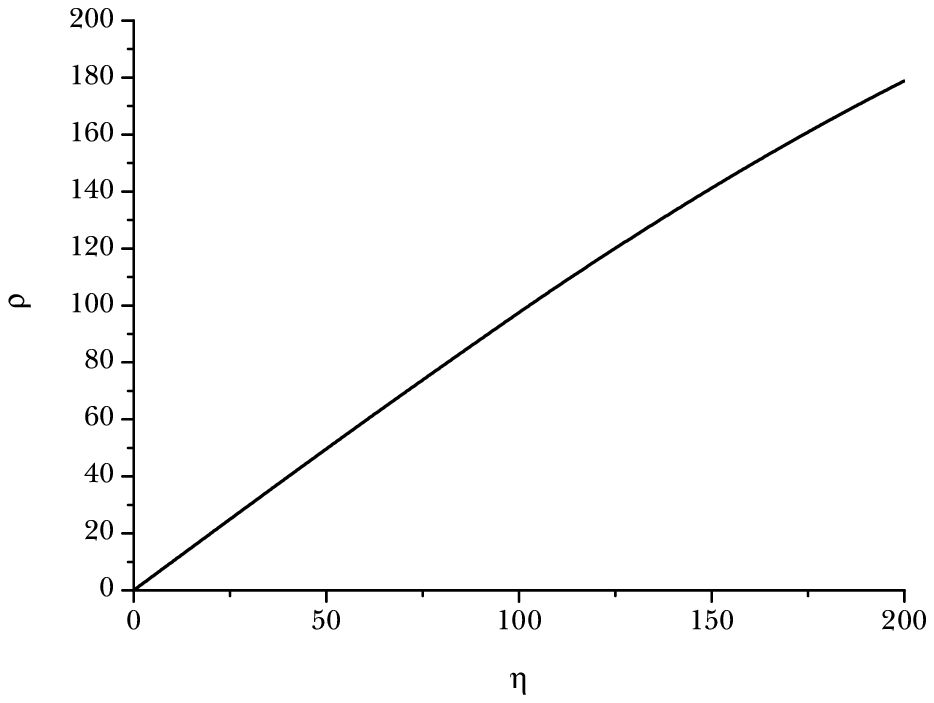}
\includegraphics[scale=0.8]{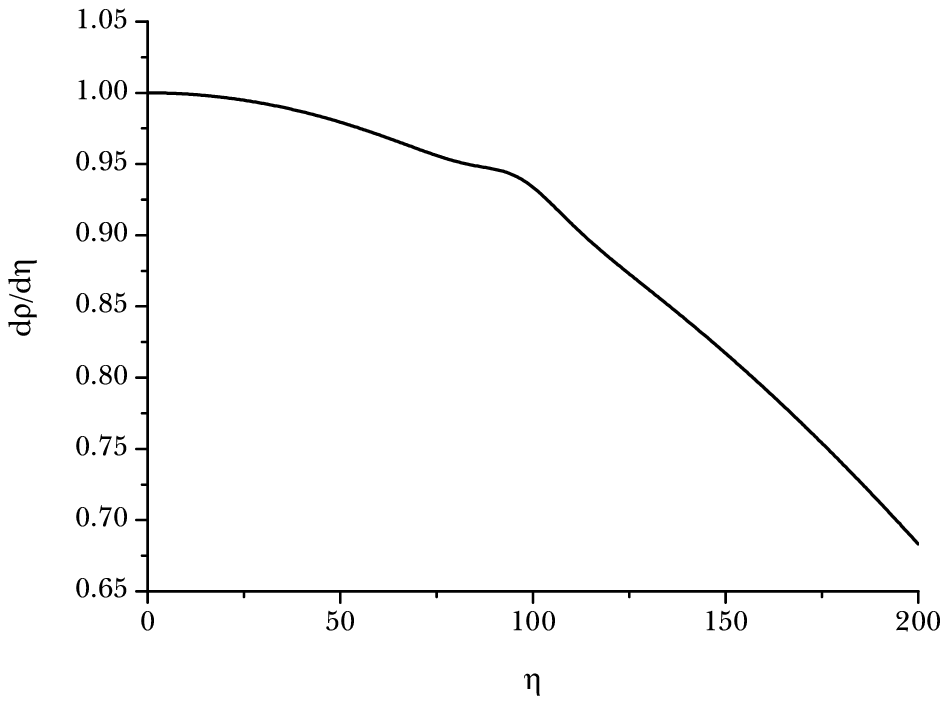}
\caption{\label{fig:type3}For $\omega = 10$, $A=10^{4}$, $\Phi_{\mathrm{f}}-\Phi_{\mathrm{t}} = 0.01$, $\delta = -0.001$, and $V_{0}=0.0001$, we obtained a false vacuum bubble solution. We plot $\Phi$, $\rho$, and $\dot{\rho}$ as functions of $\eta$.}
\end{center}
\end{figure}

We obtained the bounce solution in Figure~\ref{fig:type3}. Here, we plot $\Phi$, $\rho$, and $\dot{\rho}$. Therefore, we confirmed that a false vacuum bubble is likely to form even in the $\omega > -3/2$ cases.


\section{\label{sec:Nucandevo}Nucleation and evolution of vacuum bubbles in the thin wall approximation}

In this section, we evaluate the probability amplitude of bounces in the thin wall approximation. In the thin wall approximation, we assume that the transition region is sufficiently thin, namely,
\begin{eqnarray}
\dot{\Phi}\frac{\dot{\rho}}{\rho} \ll 1,
\end{eqnarray}
since $\dot{\Phi} \sim 0$ for the inside and the outside of the wall and $\bar{\rho}$ is sufficiently large on the wall.

We can then approximate the Einstein equation by
\begin{eqnarray}
G_{\eta\eta} &=& 3 \frac{\dot{\rho}^{2}-1}{\rho^{2}} = -3 \frac{\dot{\rho}}{\rho} \frac{\dot{\Phi}}{\Phi} + \frac{\omega}{2} \left(\frac{\dot{\Phi}}{\Phi}\right)^{2} - \frac{V}{2\Phi} \\
&\simeq& \frac{\omega}{2} \left(\frac{\dot{\Phi}}{\Phi}\right)^{2} - \frac{V}{2\Phi}.
\end{eqnarray}
Then we approximate $\dot{\rho}$ by
\begin{eqnarray}
\dot{\rho}^{2} &=& 1 - \rho \dot{\rho} \frac{\dot{\Phi}}{\Phi} + \frac{\rho^{2}}{6 \Phi^{2}} \left( \omega \dot{\Phi}^{2} - \Phi V \right)\\
&\simeq& 1 + \frac{\rho^{2}}{6 \Phi^{2}} \left( \omega \dot{\Phi}^{2} - \Phi V \right),
\end{eqnarray}
and obtain
\begin{eqnarray}
\frac{d\rho}{d\eta} = \sqrt{1 + \frac{\rho^{2}}{6 \Phi^{2}} \left( \omega \dot{\Phi}^{2} - \Phi V \right)}.
\end{eqnarray}
Therefore, inside and outside of the wall, we obtain
\begin{eqnarray}\label{eq:drhodeta}
\frac{d\rho}{d\eta} = \sqrt{1 - \frac{\rho^{2}}{6 \Phi} V}.
\end{eqnarray}

Also, we can obtain the following from the field equation by the thin wall approximation:
\begin{eqnarray}
\dot{\Phi} \ddot{\Phi} &=& \frac{1}{2} \frac{d \dot{\Phi}^{2}}{d \eta} \\
&\simeq& \left( \frac{1}{2 \omega + 3} \left(\Phi V'(\Phi) - 2V(\Phi)\right) \right) \frac{d \Phi}{d \eta}
\end{eqnarray}
and
\begin{eqnarray}
\frac{d\Phi}{d\eta}
&=& \sqrt{\frac{2}{2 \omega + 3}} \sqrt{\int_{\Phi_{\mathrm{i}}}^{\Phi}\left( \bar{\Phi} V'(\bar{\Phi}) - 2V(\bar{\Phi}) \right) d \bar{\Phi}}\\
&=& \sqrt{\frac{2}{2 \omega + 3}} \sqrt{U(\Phi)-U(\Phi_{\mathrm{i}})} \label{eq:dPhideta},
\end{eqnarray}
where $\Phi_{\mathrm{i}}$ is the field value of the inside of the bubble. Then, if $U(\Phi_{\mathrm{i}}) < U(\Phi)$ and we consider a true vacuum bubble in a false vacuum background, $2 \omega + 3$ should be positive; if $U(\Phi_{\mathrm{i}}) > U(\Phi)$ and we consider a false vacuum bubble in a true vacuum background, $2 \omega + 3$ should be negative. In our previous bounce examples, these correlations hold for all cases.

The probability amplitude is then
\begin{eqnarray}
P \sim A e^{-B},
\end{eqnarray}
where
\begin{eqnarray}
B = B_{\mathrm{outside}} + B_{\mathrm{wall}} + B_{\mathrm{inside}}
\end{eqnarray}
and
\begin{eqnarray}
B_{\mathrm{outside}} &=& S_{\mathrm{E}}(\mathrm{bounce}| \rho>\bar{\rho}) - S_{\mathrm{E}}(\mathrm{background}| \rho>\bar{\rho}),\\
B_{\mathrm{wall}} &=& S_{\mathrm{E}}(\mathrm{bounce}| \rho=\bar{\rho}) - S_{\mathrm{E}}(\mathrm{background}| \rho=\bar{\rho}),\\
B_{\mathrm{inside}} &=& S_{\mathrm{E}}(\mathrm{bounce}| \rho<\bar{\rho}) - S_{\mathrm{E}}(\mathrm{background}| \rho<\bar{\rho}).
\end{eqnarray}
Here, $S_{\mathrm{E}}(\cdots|\rho>\bar{\rho})$, $S_{\mathrm{E}}(\cdots|\rho=\bar{\rho})$, and $S_{\mathrm{E}}(\cdots|\rho<\bar{\rho})$ denote integrations of the Lagrangian density at the solution (bounce or background) for $\rho>\bar{\rho}$, $\rho=\bar{\rho}$, and $\rho<\bar{\rho}$, respectively.

\subsection{True vacuum bubbles in a false vacuum background}

For a true vacuum bubble in a false vacuum background, we demand the following field combination:
\begin{eqnarray}
\Phi(\eta) = \left\{ \begin{array}{ll}
\Phi_{\mathrm{f}} & \rho(\eta) > \bar{\rho},\\
1 & \rho(\eta) < \bar{\rho},
\end{array} \right.
\end{eqnarray}
where $\bar{\rho}$ is the location of the wall and the transition region is sufficiently thin.
Here, we assume that
\begin{eqnarray}
V(\Phi) = \left\{ \begin{array}{ll}
V_{0} & \Phi = 1,\\
\Lambda & \Phi = \Phi_{\mathrm{f}}.
\end{array} \right.
\end{eqnarray}

We then obtain the following quantities [using Equations~(\ref{eq:drhodeta}) and (\ref{eq:dPhideta})]:
\begin{eqnarray}
B_{\mathrm{outside}} &=& 0
\end{eqnarray}
\begin{eqnarray}
B_{\mathrm{wall}} &=& \frac{\pi}{4} \int d\eta \left( \bar{\rho}^{3} V(\Phi) - 6 \bar{\rho} \Phi - \bar{\rho}^{3} \Lambda + 6 \bar{\rho} \Phi_{\mathrm{f}} \right) \\
&=& \frac{\pi}{4} \sqrt{\left| \frac{2 \omega + 3}{2} \right|} \int_{1}^{\Phi_{\mathrm{f}}} \frac{d\Phi}{\sqrt{\left| U(\Phi)-U(1) \right|}} \left( \bar{\rho}^{3} V(\Phi) - 6 \bar{\rho} \Phi - \bar{\rho}^{3} \Lambda + 6 \bar{\rho} \Phi_{\mathrm{f}} \right) \\
&\equiv& 2 \pi^{2} \bar{\rho}^{3} \sigma(\omega, \bar{\rho}),
\end{eqnarray}
\begin{eqnarray}
B_{\mathrm{inside}} &=& \frac{\pi}{4} \int d \eta \left(\rho^{3} V_{0} - 6 \rho - \rho^{3} \Lambda + 6 \rho \Phi_{\mathrm{f}} \right)\\
&=& \frac{3 \pi}{2} \left[ - \frac{2}{V_{0}} \left( 1-\left( 1- \frac{V_{0}}{6}\bar{\rho}^{2} \right) ^{3/2}\right) +  \frac{2 \Phi_{\mathrm{f}}^{2}}{\Lambda} \left( 1-\left( 1- \frac{\Lambda}{6 \Phi_{\mathrm{f}}}\bar{\rho}^{2} \right) ^{3/2}\right) \right].
\end{eqnarray}

Here, we define the tension function of the thin wall $\sigma \equiv B_{\mathrm{wall}}/2 \pi^{2} \bar{\rho}^{3}$ which is a function of $\omega$ and $\bar{\rho}$.

\begin{figure}
\begin{center}
\includegraphics[scale=0.47]{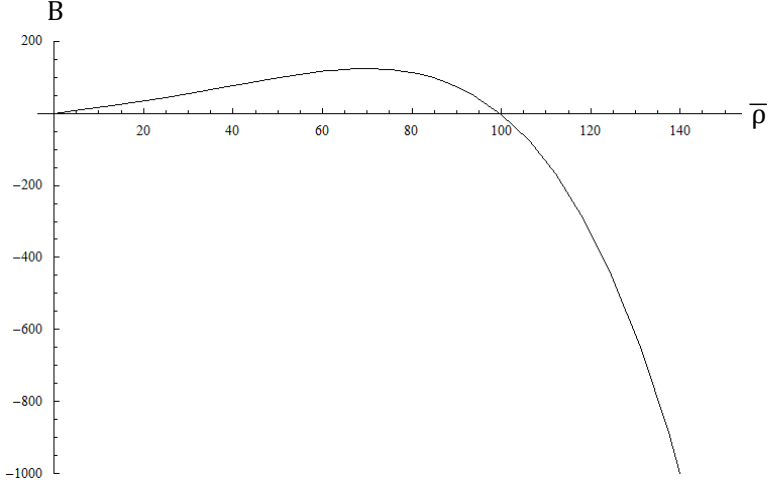}
\caption{\label{fig:B_TV}$B(\bar{\rho})$ for a true vacuum bubble in a false vacuum background: $\omega = 10$, $A=10^{4}$, $\delta = 0.0025$, $\Phi_{\mathrm{f}}-\Phi_{\mathrm{t}} = 0.025$, and $V_{0}=0$.}
\end{center}
\end{figure}

As an example, in Figure~\ref{fig:B_TV}, we plot the function $B$ for the $\omega = 10$, $A=10^{4}$, $\delta = 0.0025$, $\Phi_{\mathrm{f}}-\Phi_{\mathrm{t}} = 0.025$, and $V_{0}=0$ case. This figure shows that there is a stationary point that indicates the size of the bubble and probability.

\begin{figure}
\begin{center}
\includegraphics[scale=0.5]{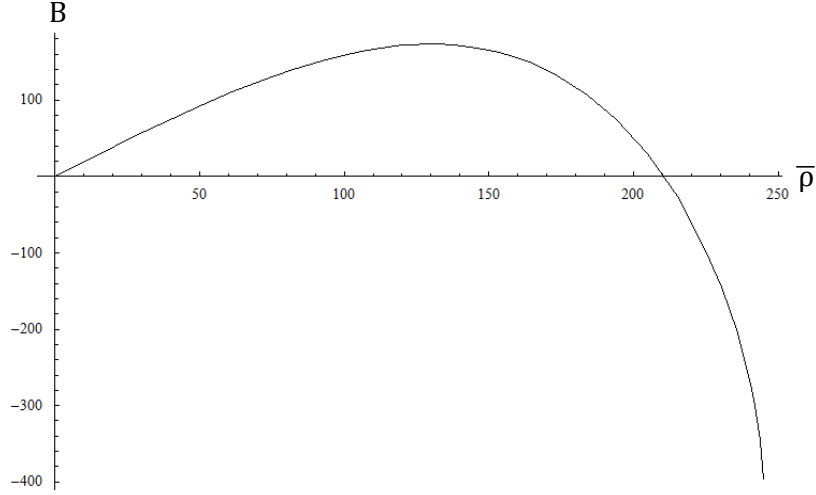}
\caption{\label{fig:B_type3}$B(\bar{\rho})$ for a false vacuum bubble in a true vacuum background: $\omega = 10$, $A=10^{4}$, $\Phi_{\mathrm{f}}-\Phi_{\mathrm{t}} = 0.01$, $\delta = -0.001$, and $V_{0}=0.0001$. Then $V(T) < V(F)$ and $U_{E}(T) > U_{E}(F)$.}
\end{center}
\end{figure}

\subsection{False vacuum bubbles in a true vacuum background}

We demand the following field combination:
\begin{eqnarray}
\Phi(\eta) = \left\{ \begin{array}{ll}
1 & \rho(\eta) > \bar{\rho},\\
\Phi_{\mathrm{f}} & \rho(\eta) < \bar{\rho},
\end{array} \right.
\end{eqnarray}
where $\bar{\rho}$ is the location of the wall and the transition region is sufficiently thin.
Here, we assume that
\begin{eqnarray}
V(\Phi) = \left\{ \begin{array}{ll}
V_{0} & \Phi = 1,\\
\Lambda & \Phi = \Phi_{\mathrm{f}}.
\end{array} \right.
\end{eqnarray}

Then, as in the previous subsection, we obtain the following quantities [using Equations~(\ref{eq:drhodeta}) and (\ref{eq:dPhideta})]:
\begin{eqnarray}
B_{\mathrm{outside}} &=& 0,
\end{eqnarray}
\begin{eqnarray}
B_{\mathrm{wall}} &=& \frac{\pi}{4} \int d\eta \left( \bar{\rho}^{3} V(\Phi) - 6 \bar{\rho} \Phi - \bar{\rho}^{3} V_{0} + 6 \bar{\rho} \right) \\
&=& \frac{\pi}{4} \sqrt{\left| \frac{2 \omega + 3}{2} \right|} \int_{\Phi_{\mathrm{f}}}^{1} \frac{d\Phi}{\sqrt{\left| U(\Phi)-U(\Phi_{\mathrm{f}}) \right|}} \left( \bar{\rho}^{3} V(\Phi) - 6 \bar{\rho} \Phi - \bar{\rho}^{3} V_{0} + 6 \bar{\rho} \right) \\
&\equiv& 2 \pi^{2} \bar{\rho}^{3} \sigma(\omega, \bar{\rho}),
\end{eqnarray}
\begin{eqnarray}
B_{\mathrm{inside}} &=& \frac{\pi}{4} \int d \eta \left(\rho^{3} \Lambda - 6 \rho \Phi_{\mathrm{f}} - \rho^{3} V_{0} + 6 \rho \right)\\
&=& \frac{3 \pi}{2} \left[ - \frac{2 \Phi_{\mathrm{f}}^{2}}{\Lambda} \left( 1-\left( 1- \frac{\Lambda}{6 \Phi_{\mathrm{f}}}\bar{\rho}^{2} \right) ^{3/2}\right) + \frac{2}{V_{0}} \left( 1-\left( 1- \frac{V_{0}}{6}\bar{\rho}^{2} \right) ^{3/2}\right) \right].
\end{eqnarray}

As an example, in Figure~\ref{fig:B_type3}, we plot the function $B$ for the $\omega = 10$, $A=10^{4}$, $\Phi_{\mathrm{f}}-\Phi_{\mathrm{t}} = 0.01$, $\delta = -0.001$, and $V_{0}=0.0001$ case. This figure shows that there is a stationary point that indicates the size of the bubble and probability.

\subsection{Dynamics of vacuum bubbles in the Lorentzian signatures}

Let us assume that a thin wall bubble has a field value $\Phi_{-}$ inside of it and $\Phi_{+}$ outside of it. The junction equation then takes the following form \cite{Lee:2010yd}:
\begin{equation}\label{eq:jc001}
\epsilon_{-} \Phi_{-}\sqrt{\dot{\bar{\rho}}^2 + f_{-}} - \epsilon_{+} \Phi_{+} \sqrt{\dot{\bar{\rho}}^2 + f_{+}} = 4\pi \bar{\rho} \sigma_{0},
\end{equation}
where
\begin{equation}
f_{\pm} = 1 - \frac{V(\Phi_{\pm})}{6\Phi_{\pm}}\bar{\rho}^{2}
\end{equation}
and $\epsilon_{\pm}$ are $+1$ if the outward normal to the wall is pointing towards increasing $\bar{\rho}$ and $-1$ if pointing towards decreasing $\bar{\rho}$.

Note that the true vacuum bubble case and the false vacuum case can be interchanged by $- \leftrightarrows +$ of each subscript and by $\sigma \rightarrow - \sigma$. Then, it is equivalent to the change $\epsilon_{\pm} \rightarrow - \epsilon_{\pm}$ and the flip $- \leftrightarrows +$ of each subscript. However, to obtain the potential $V_{\mathrm{eff}}(\bar{\rho})$ which obeys
\begin{eqnarray}
\frac{1}{2}\dot{\bar{\rho}}^{2} + V_{\mathrm{eff}}(\bar{\rho}) = 0,
\end{eqnarray}
we do not need to know the signs of each root. Therefore, the analysis of effective potentials is the same for both cases.

The authors studied the effective potential $V_{\mathrm{eff}}$ in \cite{Lee:2010yd} for false vacuum bubbles. It was realized that there are two effective potentials $V^{(1,2)}_{\mathrm{eff}}(\bar{\rho})$, and it is not difficult to confirm that each effective potential is a monotonically decreasing function. Therefore, the causal structures are determined by $\epsilon_{\pm}$ in the $\bar{\rho} \rightarrow \infty$ limit. The sign structures for $\epsilon_{\pm}$ are given in Tables~\ref{table:true} and \ref{table:false}. (We used the results in \cite{Lee:2010yd} to obtain Table~\ref{table:false}, and we flipped the signs to obtain Table~\ref{table:true}: $\epsilon_{\pm} \rightarrow - \epsilon_{\pm}$ and flip $- \leftrightarrows +$.) Also, the contents of each root never become zero; this implies that the asymptotic $\epsilon_{\pm}$ is always correct for our cases.

\begin{table}
\begin{center}
\begin{tabular}{c|c|c}
\hline
& \;\;\;\;\;\;$\Phi_{+} > 1$\;\;\;\;\;\; & \;\;\;\;\;\;$\Phi_{+} < 1$\;\;\;\;\;\;\\
\hline \hline
$\epsilon^{(1)}_{+}$ & $\pm$ & $\pm$\\
\hline
$\epsilon^{(1)}_{-}$ & $+$ & $+$\\
\hline
$\epsilon^{(2)}_{+}$ & $-$ & $+$\\
\hline
$\epsilon^{(2)}_{-}$ & $-$ & $+$\\
\hline
\end{tabular}
\caption{\label{table:true}Summary of the signs for true vacuum bubbles. The $\pm$ depends on tensions.}
\end{center}
\end{table}
\begin{table}
\begin{center}
\begin{tabular}{c|c|c}
\hline
& \;\;\;\;\;\;$\Phi_{-} > 1$\;\;\;\;\;\; & \;\;\;\;\;\;$\Phi_{-} < 1$\;\;\;\;\;\;\\
\hline \hline
$\epsilon^{(1)}_{+}$ & $-$ & $-$\\
\hline
$\epsilon^{(1)}_{-}$ & $\mp$ & $\mp$\\
\hline
$\epsilon^{(2)}_{+}$ & $+$ & $-$\\
\hline
$\epsilon^{(2)}_{-}$ & $+$ & $-$\\
\hline
\end{tabular}
\caption{\label{table:false}Summary of the signs for false vacuum bubbles. The $\mp$ depends on tensions.}
\end{center}
\end{table}

If we vary $B = B_{\mathrm{outside}} + B_{\mathrm{wall}} + B_{\mathrm{inside}}$ with respect to $\bar{\rho}$ for both the true and false vacuum bubble cases, we obtain
\begin{eqnarray}
0=\frac{\partial B}{\partial \bar{\rho}} &=& 6 \pi^{2} \bar{\rho}^{2} \sigma(\omega,\bar{\rho}) + 2 \pi^{2} \bar{\rho}^{3} \frac{\partial \sigma(\omega, \bar{\rho})}{\partial \bar{\rho}} + \frac{3 \pi}{2} \bar{\rho} \left( - \Phi_{-} \sqrt{f_{-}} + \Phi_{+} \sqrt{f_{+}} \right)\\
&=& \frac{3 \pi}{2} \bar{\rho} \left( 4 \pi \bar{\rho} \left(\frac{\bar{\rho}}{3}\frac{\partial \sigma}{\partial \bar{\rho}}+ \sigma\right) - \Phi_{-} \sqrt{f_{-}} + \Phi_{+} \sqrt{f_{+}} \right).
\end{eqnarray}
If we define $\sigma_{0} \equiv \frac{\bar{\rho}}{3}\frac{\partial \sigma}{\partial \bar{\rho}}+ \sigma$, we can derive the solution of Equation~(\ref{eq:jc001}) and $\epsilon_{\pm} = + 1$. This smoothly joins the Euclidean patch to the Lorentzian patch at the $t=0$ surface.

Here, $\epsilon_{\pm} = + 1$ implies that for true vacuum bubbles, we use $V^{(1)}_{\mathrm{eff}}$ or $V^{(2)}_{\mathrm{eff}}$ with $\Phi_{+} < 1$; $\Phi_{+} > 1$ and $V^{(2)}_{\mathrm{eff}}$ is disallowed. However, for false vacuum bubbles, only $\Phi_{-} > 1$ and $V^{(2)}_{\mathrm{eff}}$ is allowed. For each allowed case, $\epsilon_{\pm} = +1$ implies that each bubble expands over the background. For the false vacuum case, the result is consistent with our previous paper \cite{Lee:2010yd}: the only expanding bubble in a nearly flat background is for the $\Phi_{-} > 1$ and $V^{(2)}_{\mathrm{eff}}$ case.

\begin{figure}
\begin{center}
\includegraphics[scale=0.5]{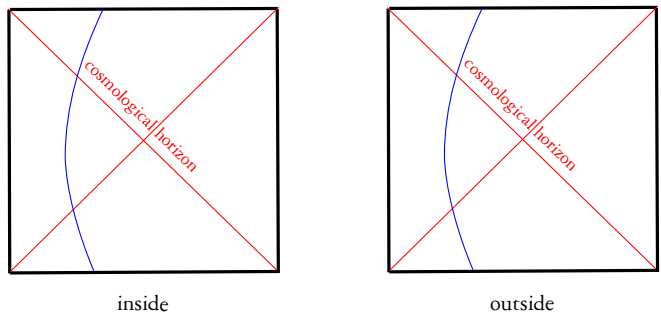}
\caption{\label{fig:thinshell}Causal structures for $\mathrm{dS}-\mathrm{dS}$ cases. For a true vacuum bubble, the inside cosmological horizon is larger than the outside cosmological horizon; for a false vacuum bubble, the inside cosmological horizon is smaller than the outside cosmological horizon.}
\end{center}
\end{figure}

For all cases, if the solution is allowed, $\epsilon_{\pm} = +1$ implies that each bubble can expand over the background. For a false vacuum bubble case, $\Phi_{-} > 1$ is not allowed and may imply that such a bubble is unstable even though it may be nucleated. For allowed false vacuum solutions, the causal structure of the wall is given by Figure~\ref{fig:thinshell}. For a false vacuum bubble, there is a time when a false vacuum bubble is larger than the inner cosmological horizon, while it is smaller than the outer cosmological horizon. In this case, if one transmits a pulse of energy to the bubble and induces an apparent horizon, and if the apparent horizon is larger than the size of the bubble, a de Sitter black hole can be seen that separates the inside bubble universe from the outside.

\section{\label{sec:Dis}Discussion}

In this paper, we explored the nucleation of vacuum bubbles in the Brans-Dicke type theory of gravity. In the Euclidean signature, we first evaluated the fields at the vacuum bubbles as solutions of the Euler-Lagrange equations of motion. Second, we calculated the bubble nucleation probabilities by integrating the Euclidean action assuming the thin wall approximation.

We illustrated three possible ways to obtain vacuum bubbles: true vacuum bubbles for $\omega > -3/2$, false vacuum bubbles for $\omega < -3/2$, and false vacuum bubbles for $\omega > -3/2$ when the vacuum energy of the false vacuum in the potential of the Einstein frame is \textit{less} than that of the true vacuum. After the bubble is nucleated at the $t=0$ surface, we can smoothly interpolate the field combinations to some solutions of the Lorentzian signature and consistently continue their subsequent evolutions.

Next, it might be relevant to address the issue of the conformal frame choice between the Jordan frame and the Einstein frame.
In the previous work \cite{Lee:2010yd}, we studied dynamics of expanding small false vacuum bubbles in the Brans-Dicke theory by using the thin wall approximation. The effect of the non-minimal coupling of the Brans-Dicke field makes the effective tension of the wall negative, and therefore the small false vacuum bubble can expand to the surrounding background.
If there is an expanding small false vacuum bubble, it will violate the null energy condition around the wall when it begins to inflate \cite{Lee:2010yd}\cite{Hansen:2009kn}. If we consider this fact, for false vacuum bubbles with $\omega < -3/2$, the nucleation of the bubbles is not strange since $\omega < -3/2$ means that the conformal transformation is not well defined or the defined scalar field in the Einstein frame behaves as a ghost since its kinetic term in the Lagrangian fails to be positive definite and becomes negative. However, if a false vacuum bubble for $\omega > -3/2$ is possible, it seems to be strange. In the Jordan frame, this is not a problem since working in the Jordan frame can violate the null energy condition \cite{Kang:1996rj}; however, working in the Einstein frame with $\omega > -3/2$ does not violate the null energy condition and it would be a problem. In the present work, however, our solution does not suffer from this paradoxical situation, since a false vacuum bubble in the Jordan frame corresponds to a true vacuum bubble in the Einstein frame. In the Einstein frame, therefore, there is no reason to conclude that such a bubble violates the null energy condition. Also, it is not so strange, although we obtain a small false vacuum bubble, since the nucleation of a true vacuum bubble in the Einstein frame is generally possible. \footnote{In Einstein gravity, the nucleation of a large false vacuum bubble is possible in the de Sitter space \cite{LW}. The possible types in de Sitter space have been studied in \cite{LL}. The relation of vacuum bubble solutions between the Jordan frame and the Einstein frame should be clarified, and we open this issue for future work, whether it corresponds to a true/false vacuum bubble or to a small/large vacuum bubble.}

Then, two interesting questions arise. First, could we define and perform the conformal transformation on the wall? Second, which conformal frame would be physical? Indeed, this is a long-standing problem which has not been resolved yet.
In addition, for the gravity theory with only tensor degrees of freedom, the choice of relevant conformal frame via conformal transformations is the usual issue to address. For the scalar-tensor gravity theories such as the present Brans-Dicke type theory or the superstring theories, however, the conformal transformation needs to be extended to the Weyl rescaling.
Therefore, when one investigates the physical characteristics of a conformal frame, the Weyl rescaling of the scalar field should be carefully taken into account as well for a consistent study. In the present work, however, this issue has not been properly addressed but we expect that even if we consider the Weyl rescaling of the scalar degree of freedom, the conclusions presented in the present work would remain essentially the same.

If the Einstein frame is physical, our conclusions in Section~\ref{sec:via} will be irrelevant, and further studies of the nucleation of small false vacuum bubbles should be done. However, if the Jordan frame were physical, the generation of a small false vacuum bubble would be allowed. It is known that there is controversy regarding the latter question \cite{Faraoni:2004pi}, and the most conservative interpretation is that the two frames are equivalent, at least at the classical level. If we include quantum effects in a given frame, then the frame would be physical. In this sense, the choice of the Jordan frame is still a viable option. We leave these two questions open until they can be answered in the future work.

We now conclude that, in general, the scalar-tensor theories or the Brans-Dicke type theories, which may include and represent certain features of string theory, allow vacuum bubble solutions, not only true vacuum bubbles but also false vacuum bubbles. One potential problem is that, if we wish to derive a scalar-tensor or Brans-Dicke type model from string theory, it would be in dilaton gravity; but in dilaton gravity, the potential of the dilaton field would be restricted by the theory. If we assume a special condition on the potential, then dilaton gravity could generate a false vacuum bubble; however, it is still unclear whether dilaton gravity can indeed admit such a special potential. Therefore, we also leave this question open for future work.

\section*{Acknowledgments}

DY would like to thank Ewan Stewart for discussions and encouragement.
DY and YJL were supported by Korea Research Foundation Grants  No. KRF-313-2007-C00164 and No. KRF-341-2007-C00010 funded by the Korean government (MOEHRD) and BK21.
DY, BHL and WL were supported by a Korea Science and Engineering Foundation (KOSEF) grant funded by the Korean government (MEST) through the Center for Quantum Spacetime (CQUeST) of Sogang University with Grant No. R11 - 2005 - 021.
WL was supported by the National Research Foundation of Korea funded by the Korean Government (Ministry of Education, Science and Technology) Grant No. NRF-2010-355-C00017.
HK was supported by the research fund of the KVN division at KASI.

\end{document}